\pdfoutput=1

\documentclass[%
 reprint,
superscriptaddress,
 amsmath,amssymb,
 aps,
prb,
floatfix
]{revtex4-1}

\usepackage{graphicx}
\usepackage{dcolumn}
\usepackage{bm}
\usepackage{color}

\newcommand{\bvec}[1]{\ensuremath{\mathbf{#1}}}
\newcommand{\bsym}[1]{\ensuremath{\boldsymbol{#1}}}

\begin{document}

\author{F\'elix Mouhat}
\affiliation{IMPMC, Sorbonne Universit\'es, UPMC Univ Paris 06, CNRS, IRD, MNHN, 4 place Jussieu, 75252 Paris, France}
 \email{felix.mouhat@impmc.upmc.fr}
\author{Sandro Sorella}
\affiliation{International School for Advanced Studies (SISSA), Via Bonomea 26, 34136 Trieste, Italy and INFM Democritos National Simulation Center, 34151 Trieste, Italy}
\author{Rodolphe Vuilleumier}
\affiliation{\'Ecole normale sup\'erieure, PSL Research University, UPMC Univ Paris 06, CNRS, D\'epartement de Chimie, PASTEUR, 24, rue Lhomond, 75005 Paris, France}
\affiliation{Sorbonne Universit\'es, UPMC Univ Paris 06, ENS, CNRS, PASTEUR, 75005 Paris, France}
\author{Antonino Marco Saitta}
\affiliation{IMPMC, Sorbonne Universit\'es, UPMC Univ Paris 06, CNRS, IRD, MNHN, 4 place Jussieu, 75252 Paris, France}
\author{Michele Casula}
\affiliation{IMPMC, Sorbonne Universit\'es, UPMC Univ Paris 06, CNRS, IRD, MNHN, 4 place Jussieu, 75252 Paris, France}

\title{Fully quantum description of the Zundel ion: combining variational quantum Monte Carlo with path integral Langevin dynamics}



\begin{abstract}
We introduce a novel approach 
for a fully quantum description of coupled electron-ion systems from first principles.
It combines the variational quantum Monte Carlo solution of the electronic part with the path integral formalism for the quantum nuclear dynamics. 
On the one hand, the path integral molecular dynamics 
includes nuclear quantum effects 
by adding a set of fictitious classical particles (beads) aimed at reproducing nuclear quantum fluctuations via a harmonic kinetic term.
On the other hand, variational quantum Monte Carlo can provide 
Born-Oppenheimer potential energy surfaces 
with a precision comparable to the most advanced post Hartree-Fock approaches, and with a favorable scaling with the system size. 
In order to deal with the intrinsic noise due to the stochastic nature of quantum Monte Carlo methods, we generalize the path integral 
molecular dynamics
using a Langevin thermostat correlated according to the covariance matrix of quantum Monte Carlo nuclear forces. The variational parameters of the quantum Monte Carlo wave function are evolved during the nuclear dynamics, such that the Born-Oppenheimer potential energy surface is unbiased. Statistical errors on the wave function parameters are reduced by resorting to bead grouping average, which we show to be accurate and well controlled.
Our general algorithm relies on a Trotter breakup between the dynamics driven by ionic forces and the one set by the harmonic interbead couplings. The latter is exactly integrated even in presence of the Langevin thermostat, thanks to the mapping onto an Ornstein-Uhlenbeck process. 
This framework turns out to be very efficient also in the case of noiseless (deterministic) ionic forces.
The new implementation is validated on the Zundel ion ($\text{H}_5\text{O}_2^+$) by direct comparison with standard path integral Langevin dynamics calculations made with a coupled cluster potential energy surface. Nuclear quantum effects are confirmed to be dominant over thermal effects well beyond room temperature giving the excess proton an increased mobility by quantum tunneling.     
\end{abstract}

\maketitle

\section{\label{intro}Introduction}

In spite of about half a century of numerical investigations aimed at providing a quantitative and reliable microscopic description of liquid water, there is still a difficulty 
in describing
accurately proton hopping inside the strong and well structured hydrogen bond network giving water its numerous peculiar features. Since water is the solvent of most chemical and biological processes occurring on Earth, one of the long-term goals is to explicitly take into account, at a high accuracy level, hydrogen-bond driven phenomena in the treatment of species interacting with this solvent. The solvation and transport properties of hydronium ($\text{H}_3\text{O}^+$) and hydroxide (HO$^-$) ions in aqueous environment have enormous impacts on areas ranging from aqueous acid-base chemistry\cite{PTaqueous_sol1,PTaqueous_sol2,PTaqueous_sol3,Hassanali2011,Lee2011,Agmon2016}, enzymatic proton transfer\cite{Riccardi2006,Riccardi2010}, as well as proton transfer in biological channels\cite{Wraight2006}, through fuel cell membranes\cite{phys_rev,pnas_membrane} and on ice surfaces facilitating atmospheric reactions\cite{Gertner1998}. Recent progress in our understanding of proton transport in water has been reviewed in Refs.~\onlinecite{marx2006} and \onlinecite{Hassanali2014}. 

Due to a variety of interactions present in water such as weak and dispersive dipolar interactions that compete during the formation and the breaking of hydrogen bonds inside a strongly polar network, an accurate description of the Potential Energy Surface (PES) is required to fully understand the microscopic properties of water. From a theoretical point of view, liquid water has been studied through Molecular Dynamics (MD) techniques such as the empirical force fields methods\cite{force_field1, force_field2,force_field3} and \emph{ab initio} molecular dynamics (AIMD)\cite{water1_parri,water2,water3,water4,abinitio_bio1,abinitio_bio2}. Because of its good scalability with the system size which is suited to study liquids, Density Functional Theory (DFT) has been the most employed technique to perform AIMD simulations of bulk water. Nevertheless, DFT-based calculations performed with Generalized Gradient Approximation (GGA) functionals give rise to an overstructuration of water: melting point temperature\cite{water_temperature} and oxygen-oxygen radial distribution function\cite{water_radial} are for instance poorly reproduced. Some improvements by including Van der Waals corrections such as the BLYP-DCACP, BLYP-D$_{G2}$ or PBE0+TS-vdW(SC) functionals have been tested and proven to get results closer to experiments\cite{Lin2012,Santra2015} but these approaches suffer from a lack of generality and transferability. Gillan \emph{et al.} recently established a numerical scoring system to evaluate the quality of different exchange-correlation functionals for a wide range of water systems\cite{Gillan2016}.

As an alternative to DFT-AIMD, very accurate calculations using embedded-fragment second-order 
M{\o}ller-Plesset
perturbation theory (MP2) have been carried out and gave satisfactory results despite the shortness of the MD trajectories (6 ps of equilibration and 5 ps of production run)\cite{Willow2015}. Perturbative theories such as MP2/MP4 or the most advanced quantum chemistry techniques such as Coupled Cluster (CC) calculations are indeed suffering from a poor scalability with the number of particles $N$ in the system. 

To overcome this major limitation, many-body representations of water and other aqueous systems appear to be a possible route to describe accurately proton transfer properties at a cost marginally higher than that associated with empirical force fields. The many-body approach has witnessed tremendous recent advances since numerous polarizable potentials  have been proven to be closely accurate to CCSD(T) reference.\cite{Cisneros2016} Some of them, namely the MB-pol \cite{Babin2013,Babin2014,Medders2014}, provide a correct description of water from the gas to condensed phase, outperforming all models currently available. Building upon 
this formalism, 
several many-body representations have also been introduced for protonated water.\cite{Pinski2014,Yu2016} 

While keeping an explicit description of the electronic structure, Quantum Monte Carlo (QMC) techniques constitute a very interesting alternative to study aqueous systems, since the scalability of the method ($N^{3-4}$) makes the simulations of very large systems computationally affordable with a much greater accuracy than DFT in most cases. Recently, it has been demonstrated that QMC techniques provide results as accurate as the basis set converged CCSD(T) for small neutral or charged water clusters\cite{alfe,water_Needs}. Consequently, QMC can now be used as a benchmark method and 
certainly benefits from its intrinsically parallel formulation in modern supercomputers. The price to pay here is the systematic statistical incertitude arising from the stochastic nature of the QMC approach,  which can however be reduced by generating longer Markov chains as the error is inversely proportional to the square root of the number of QMC generations. 

Some strategies have thus been developed to incorporate the intrinsic QMC noise into a classical 
statistical mechanics framework
of nuclei
at finite temperature $T$. For instance, the Coupled Electron Ion Monte Carlo (CEIMC) method relies on the Born-Oppenheimer (BO) approximation for treating finite temperature ions coupled with ground state electrons. 
The Boltzmann distribution function of the ionic degrees of freedom is 
sampled at fixed temperature via a Metropolis MC simulation
based on the electronic energies computed during independent ground state QMC calculations\cite{Pierleoni2004,Pierleoni2006}. Another 
strategy is 
to resort to a Langevin MD approach
to correlate the noise driving the Langevin Dynamics (LD) of the system by the QMC forces covariance. In that case, the wave function parameters are optimized by Variational Monte Carlo (VMC) calculations along the MD path such that the electronic solution has always energies and forces as close as possible to the true BO surface\cite{Attaccalite2008,Mazzola2014}.

Last year, Zen and coworkers applied the latter method to perform the very first QMC-based MD simulation of bulk liquid water at ambient conditions using 32 water molecules in a cubic cell with periodic boundary conditions\cite{Zen2015}. By obtaining a better position of the first peak of the oxygen-oxygen radial distribution function compared with most advanced functionals, they demonstrated the ability of QMC to tackle this kind of problems. However, the authors had to make some approximations: they expanded the wave function over a small basis set and, more importantly, a classical description of the nuclei was adopted.

Indeed, it is well known that
Nuclear Quantum Effects (NQE) play a crucial role in the description of water or ice by deeply affecting the radial distribution functions and distorting hydrogen bonds because of quantum disorder\cite{Morrone2008,Vega2010,Pamuk2012,Kong2012,Giberti2014,Lobaugh1996,Markovitch2008}. In this paper, we propose to extend Zen's pioneering work by including a 
nuclear quantum description 
within the QMC-driven dynamics. This is achieved within a Path Integral Langevin Dynamics (PILD) approach which, to the best of our knowledge, has never been used in the case of non-deterministic forces. 

On the one hand, we derive an original algorithm to integrate the Langevin equations of motion for quantum particles. The designed algorithm is very general since it can be used to propagate both deterministic and stochastic forces. On the other hand, within our framework we show that the inclusion of the quantum nature of the nuclei in the QMC-driven MD does not slow down the simulation with respect to the classical counterpart. This is thanks to an effective average of wave function parameters in the polymer representation of quantum nuclei, which relies upon a bead-grouping technique. Therefore, the cost of the wave function optimization is not increased by the inclusion of the additional quantum degrees of freedom. 

To test our methodological development, we apply it to $\text{H}_5\text{O}_2^+$, namely the Zundel ion\cite{zundel_model}, widely used as benchmark system. Indeed, it is the simplest system to exhibit a non trivial proton transfer and its reduced size makes a comprehensive and systematic study of the problem easier.
More importantly, there is a huge amount of data\cite{DFT_Zundel,parri_science,rodolphe,Vuilleumier1998,Vuilleumier2000,brueckner,aaauer,wales,extensive_H5O2,zundel_vibrational,Park2007} to compare our results with, because the description of excess proton in water has been massively studied both theoretically and experimentally in the last fifty years. We can cite for instance the extremely accurate Potential Energy Surface (PES) generated by Bowman and coworkers from almost exact CCSD(T) calculations\cite{huang}, the MS-EVB method\cite{rodolphe,Vuilleumier1998,Schmitt1998,Day2002}, 
or more recently the LEWIS model developed by Herzfeld.\cite{Kale2012} Moreover, because of the great importance of the NQE in the Zundel cation, the latter is almost always used to test and validate new approaches\cite{Kapil2016}. 

The paper is organized as follows. In Section \ref{met}, we introduce the theoretical framework by first reviewing the algorithms for classical Langevin simulations with QMC forces (Sec.~\ref{class_corr}). In Sec.~\ref{quant_best} we extend the classical description of nuclei by introducing an original and efficient algorithm to integrate the equations of motion in the nuclear quantum case with or without noisy forces. A study of the numerical stability of the new path integral integrator is done in Sec.~\ref{num} for deterministic forces, where a direct comparison with existing PILD algorithms is possible. The inclusion of QMC forces is discussed in Sec.~\ref{QMC}. The general form of the QMC wave function for the electronic part is described in Sec.~\ref{QMC-wf}, the bead-grouping approximation is introduced in Sec.~\ref{QMC-bead-grouping}, while the QMC-noise correction scheme is detailed in Sec.~\ref{QMC-noise-correction}, which yields an unbiased sampling of the canonical quantum partition function even in the presence of stochastic errors. Sec.~\ref{QMC-stability} reports on the numerical stability analysis of the new integrators with noisy QMC forces.
In Section \ref{results}, we show the results obtained for the Zundel ion at low and room temperatures by taking into account, or not, the NQE and demonstrate the excellent agreement between the noisy QMC PES and the very accurate deterministic CCSD(T) one, which validates our method. Finally, in Section \ref{conclu}, we provide some concluding remarks and we discuss the possible future applications of the algorithms developed in this paper.

\section{\label{met}Methods}

In this Section, we derive the methodology to sample the canonical partition function corresponding to Hamiltonians including either classical or quantum nuclei. This is achieved using the generalized Langevin dynamics, which has been demonstrated to be compatible with the path integral approach to include a quantum description of the nuclei\cite{Ceriotti2011}. Thanks to the stochastic nature of this category of non-Newtonian dynamics, the Langevin formulation of the equations of motion can be extended to the QMC case, where the forces are intrinsically noisy\cite{Attaccalite2008,Mazzola2014}.

\subsection{\label{class_corr}Classical Langevin Dynamics: Classical Momentum-Position Correlator (CMPC) algorithm}

Let us consider a system constituted by $N$ classical particles described by the Hamiltonian
\begin{equation}
H = \sum\limits_{i=1}^{3N} \frac{p_i^2}{2m_i} + V\left(q_1,...,q_{3N}\right),
\end{equation}
where $m_i$ are the physical masses, $p_i$ and $q_i$ are the atomic momenta and positions, respectively, and $V$ is the PES in which the system evolves. 
Hereafter, we are going to use transformed variables, more convenient to manipulate, achieved by the following mass scaling:
\begin{eqnarray}
 q_i &=& q^0_i  \sqrt{m_i}  \nonumber \\
 p_i &=& p^0_i /\sqrt{m_i}  \nonumber \\
 \eta_i &=& \eta^0_i \sqrt{m_i}  \nonumber \\
 f_i &=& f^0_i /\sqrt{m_i},
\label{mass_scaling}
\end{eqnarray}
where $\left\{q_i^0,p_i^0,\eta_i^0,f_i^0\right\}_{i=1,..,3N}$ are the original coordinates and $\left\{q_i,p_i,\eta_i,f_i\right\}_{i=1,..,3N}$ denote the transformed ones. 
After applying the above transformation, the corresponding second order Langevin dynamics reads:
\begin{eqnarray} 
 \dot {\bvec{p}}  &=&  -  \boldsymbol{\gamma}  \bvec{p} 
   +   \bvec{f}(\bvec{q}) + \bsym{\eta}(t) 
\label{vel_lang} \\ 
 \dot {\bvec{q}}   &=& \bvec{p},  
\label{pos_lang}  
\end{eqnarray} 
where $\bvec{f} = - \bsym{\nabla}_{\bvec{q}} V (\bvec{q})$ is the applied force deriving from the PES, and the vectors are $3N$-dimensional. To ensure that the equilibrium distribution generated by this dynamics is canonical, the r.h.s of Eq.~(\ref{vel_lang}) includes a $3N \times 3N$ friction matrix $\bsym{\gamma}$ and a random force vector $\bsym{\eta}$. The latter controls the temperature $T$ of the system via the Fluctuation-Dissipation Theorem (FDT)\cite{Kubo}
\begin{equation} \label{fdcond}
\boldsymbol{\gamma}= \frac{\boldsymbol{\alpha }} { 2 k_B T },
\end{equation}
with $k_B$ the Boltzmann constant, and 
\begin{equation}
\alpha_{ij}\delta(t-t^\prime) = \langle \eta_i(t) \eta_j(t^\prime)  \rangle 
\label{local_corr}
\end{equation}
is the force covariance matrix which simply reduces to a diagonal form for white noise\cite{Fixman1986,Izaguirre2001}. The dynamics generated by the stochastic differential Eqs.~(\ref{vel_lang}) and (\ref{pos_lang}) is Markovian, as the noise fluctuations are locally correlated in time. In the case the ionic forces are computed by QMC methods (as we will see in Sec.~\ref{QMC-noise-correction}), the covariance matrix has non-trivial off-diagonal elements, and $\bsym{\eta}$ fluctuations are correlated by the QMC statistics. Thus, the Langevin approach can naturally deal with noisy QMC forces, as already shown by Sorella and coworkers in the classical MD framework\cite{Attaccalite2008,Mazzola2014,Zen2015}. 

The basic integration scheme developed by Attacalite \emph{et al.}\cite{Attaccalite2008} to perform QMC-driven MD simulations at finite temperature
has been further developed in Refs.~\onlinecite{Mazzola2014} and \onlinecite{Zen2015}. In the latter scheme, dubbed as Classical Momentum-Position Correlator (CMPC) algorithm, the Langevin dynamics is driven by a correlated noise affecting both momenta and positions. To set up the formalism that we will be exploiting in the quantum path integral case, we provide here an alternative derivation, based on the joint momentum-position coordinates
\begin{equation}
 \bvec{X} = \begin{pmatrix}
 \bvec{p} \\
 \bvec{q} 
\end{pmatrix},
\label{generalized_coord}
\end{equation}
where $\bvec{X}$ is a $6N$-dimensional vector. Analogously, we extend the definition of the ionic and random force vectors to be
$\bvec{F}=\begin{pmatrix}
\bvec{f} \\
\bvec{0}
\end{pmatrix}$ and $\bvec{E}=\begin{pmatrix}
\bsym{\eta} \\
\bvec{0}
\end{pmatrix}$, respectively.
In this extended basis, Eqs.~(\ref{vel_lang}) and (\ref{pos_lang}) can be rewritten into a generalized SDE, which reads as
\begin{equation}
 \dot {\mathbf{X}} = -\hat {\boldsymbol{\gamma}} \mathbf{X} + \bvec{F} + \bvec{E},   
\label{ext_class_lang}
\end{equation}
where in this notation the matrix $\hat {\boldsymbol{\gamma}}$ represents a generalized friction that couples 
both momenta and positions by:
\begin{equation}
\hat {\boldsymbol{\gamma}} = \begin{pmatrix}
  \boldsymbol{\gamma} & \mathbf{0} \\
 -\mathbf{I} & \mathbf{0}  
\end{pmatrix},
\end{equation}
with the ``physical'' friction $\boldsymbol \gamma$ being the same $3N \times 3N$ matrix introduced in Eq.~(\ref{vel_lang}), and $\mathbf{I}$ is the identity matrix. 
The formal solution of Eq.~(\ref{ext_class_lang}) is provided by

\footnotesize
\begin{equation}
\bvec{X}(t^\prime)= e^{-\bsym{\hat \gamma} (t^\prime- t)} \bvec{X}(t) + \int_t^{t^\prime}\!\!\!  \textrm{d}s \,  e^{\bsym{\hat \gamma}(s-t^\prime)} \left(\bvec{F}(\bvec{X}(s))+\bvec{E}(s)\right).
\label{sol_ext_class_lang}
\end{equation}
\normalsize

\noindent Using  joint
coordinates would be of little use, if we were not be able to evaluate the exponential $e^{-\bsym{\hat \gamma} \delta t}$ in a closed analytic form.
This is possible
 because the block matrix $\boldsymbol{\hat \gamma}$ can be more conveniently rewritten in terms of Pauli matrices $\boldsymbol {\sigma}_x$, $\boldsymbol {\sigma}_y$, $\boldsymbol {\sigma}_z$, as follows:
\begin{equation}
 \boldsymbol{\hat \gamma} = \frac{\bsym{\gamma}}{2} \otimes \mathbf I -\frac{\mathbf I}{2} \otimes \boldsymbol{\sigma}_x + 
i \frac{\mathbf I}{2} \otimes \boldsymbol{\sigma}_y + \frac{\bsym{\gamma}}{2} \otimes \boldsymbol{\sigma}_z.
\label{pauli_expansion}
\end{equation}
Then, the exponentiation can be straightforwardly obtained by using standard Pauli matrices algebra, and
the solution can be given in a closed form:
\begin{eqnarray} 
 \bvec{p}_{n+1}  &=&e^{ -  \boldsymbol{\gamma} \delta t  }   \bvec{p}_{n} +
\boldsymbol{ \Gamma}  \left(\bvec{f}_n +   \bsym{ \tilde  \eta} \right)  \label{sol_vel_ext} \\
\bvec{q}_{n+1} & =& \bvec{q}_n+   \boldsymbol{\Gamma} \bvec{p}_n  
+ \boldsymbol{ \Theta }\left( \bvec{f}_n +  \bsym {\tilde {\tilde \eta}}  \right),  \label{sol_pos_ext}
\end{eqnarray}
where the time evolution has been discretized with time step $\delta t$, and the subscripts refer to the corresponding time slice, such that $\bvec{p}_n = \bvec{p}(t_n)$, $\bvec{q}_n= \bvec{q}(t_n)$, and $\bvec{f}_n=\bvec{f}(\bvec{q}(t_n))$. In the above Equations, the other symbols are defined as
\begin{eqnarray}
\bsym{\Gamma} & = & \bsym{\gamma}^{-1} ( 1 - e^{ -\bsym{\gamma} \delta t } ), \label{capital_gamma} \\
\boldsymbol \Theta &= & \boldsymbol \gamma^{-2} (  -1+ \boldsymbol \gamma \delta t  +e^{-\boldsymbol \gamma \delta t} ), \label{capital_theta} \\
  \bsym{\tilde \eta}  &=& \boldsymbol \Gamma^{-1}  \int\limits_{t_n}^{t_{n+1}} dt e^{ \boldsymbol \gamma (t-t_{n+1})}  \bsym{\eta} (t),   \label{eta1} \\ 
  \bsym{\tilde { \tilde  {\eta}}}  &=&  (\boldsymbol {\Theta \gamma})^{-1}  
\int\limits_{t_n}^{t_{n+1}} dt  ( 1-  e^{ \boldsymbol \gamma (t-t_{n+1})}) \bsym{\eta} (t).  \label{eta2}
\end{eqnarray}
We immediately notice that the CMPC algorithm differs from those previously developed in literature\cite{SKEEL2002,RICCI2003,VandenEijnden2006,Bussi2007,Melchionna2007,GrnbechJensen2013}, since momenta and positions are propagated simultaneously in a single iteration thanks to the use of momentum-position correlation matrices. In particular, according to Eqs.~(\ref{sol_vel_ext}) and (\ref{sol_pos_ext}), not only the momenta but also the positions are affected by the integrated Langevin noise, $\bsym{\tilde \eta}$ and $\bsym{\tilde { \tilde  {\eta}}}$ whose properties are detailed in Appendix \ref{cmpc_correlators}. 

It is interesting to note that the momentum-position correlator formalism correlates the integrated noise, even without dealing with non-diagonal $\bsym{\alpha}$ matrices, as we will do in the QMC case (Sec.~\ref{QMC-noise-correction}). We also remark that in order to derive the integrated equations of motion (\ref{sol_vel_ext}) and (\ref{sol_pos_ext}), 
we disregarded the $\bvec{q}$-time dependence of $\bvec{f}$ in Eq.~(\ref{sol_ext_class_lang}) in the time interval $\delta t$. For deterministic forces, where $\bsym{\alpha}$ is position independent, it is the only approximation left in the discretized classical Langevin dynamics driven by Eqs. (\ref{sol_vel_ext}) and (\ref{sol_pos_ext}).

For deterministic $\bvec{f}$, $\gamma_{ij}=\gamma_\textrm{BO} \delta_{ij}$, while the definition of $\bsym{\gamma}$ is more general for noisy QMC forces (see Sec.~\ref{QMC}). Once $\bsym{\gamma}$ and $\delta t$ have been set, the numerical evolution is performed according to Eqs.~(\ref{sol_vel_ext}) and (\ref{sol_pos_ext}) in the frame which diagonalizes $\bsym{\gamma}$.

\subsection{\label{quant_best}Path Integral Ornstein-Uhlenbeck Dynamics (PIOUD)}
\label{PIOUD}

As we mentioned in the Introduction, it is well known that NQE are extremely important, in liquid water and aqueous species such as the Zundel ion, because of the light mass of the hydrogen atom. 
Several approaches to deal with NQE have been proposed so far, such as the Quantum Thermal Bath (QTB)\cite{Dammak2009,Ceriotti2009_2,Brieuc2016,Brieuc2016b}, but we have chosen here to focus on the Path Integral (PI) framework for the following reasons. First, the formulation of the theory is intrinsically parallel and perfectly compatible with the QMC multi-walker implementation. 
Second, the PI relies on approximations which are systematically improvable and the exact result can be recovered with controlled extrapolations. Third, this approach can be easily combined with the LD leading to the PILD method, for which very efficient algorithms already exist\cite{Ceriotti2010,Liu2016}. 

We first recall the basic formalism of the PIMD method (see Ref.~\onlinecite{Ceriotti2010} for further details), before introducing our approach to propagate the corresponding equations of motion. In PI theory, the quantum partition function $Z = \text{Tr}\left[e^{-\beta H}\right]$ (with $\beta = 1/k_BT$) is usually approximated after a standard Trotter factorization\cite{Trotter1959} by 
\begin{equation}
Z \simeq \frac{1}{(2\pi \hbar)^f}\int \textrm{d}^f \mathbf{p} \,\textrm{d}^f \mathbf{q} \, e^{-\tau H_M(\mathbf{p},\mathbf{q})},
\label{Z}
\end{equation}
with $f=MN$ and $\tau=\beta/M$. We have replaced here the true quantum particles by fictitious classical ring polymers consisting of $M$ replicas (beads) of the system connected to each other by harmonic springs evolving in the
imaginary time $\tau$. Without loss of generality, we can extend the definition of classical vectors to the quantum case, by including all $M$ replicas in the string, such that $\bvec{q} \equiv \left( \bvec{q}^{(1)},\ldots,\bvec{q}^{(i)},\ldots,\bvec{q}^{(M)} \right)^T$ and it becomes $3NM$-dimensional. Bold vectors will automatically include all replicas in the quantum case. If interpreted classically, the partition function $Z$ in Eq.~(\ref{Z}) describes a system at the effective temperature $M T$. The Hamiltonian $H_M$ corresponding to this quantum-to-classical isomorphism reads as

\footnotesize
\begin{eqnarray}
H_M(\mathbf{p},\mathbf{q})&=& \sum \limits_{i=1}^{3N}\sum \limits_{j=1}^M\left( \frac{1}{2} [p_i^{(j)}]^2 + \frac{1}{2}\tilde{\omega}_M^2\left(q_i^{(j)}-q_i^{(j-1)}\right)^2\right) \nonumber \\
&+& \sum \limits_{j=1}^M V(q_1^{(j)},..,q_{3N}^{(j)}),
\label{mapping}
\end{eqnarray}
\normalsize

\noindent written in mass-scaled coordinates (\ref{mass_scaling}), with $\tilde{\omega}_M = M/\beta \hbar$, and subjected to the ring boundary condition $q_i^{(0)} \equiv q_i^{(M)}$. Thus, the PIMD method is a way to calculate quantum mechanical properties using the recipes from classical statistical mechanics with a modified Hamiltonian containing an additional quantum kinetic term.

Various choices are available for the thermostat, ranging from the more conventional Nos\'e-Hoover chain\cite{Nose1984,Hoover1985} to stochastic thermostatting in both its simple path integral Langevin equation (PILE) form\cite{Ceriotti2010}, which fulfills the FDT, and its Generalized Langevin Equation (GLE) flavors\cite{Ceriotti2010,Ceriotti2010_2,Ceriotti2011}, where the noise is colored to simulate quantum effects and accelerate the convergence of the PIMD with the number of beads, at the price of breaking the validity of the FDT. Here, we keep fulfilling the FDT. However, instead of using the PILE integration algorithm\cite{Ceriotti2010}, we will extend the CMPC algorithm, described in Sec.~\ref{class_corr}, to the quantum case. As already mentioned, this framework will allow us to incorporate noisy QMC forces in the equations of motion, by using an appropriately tailored Langevin noise without breaking the FDT, as it is explained in Sec.~\ref{QMC}. Moreover, in our quantum algorithm, which makes use of a Trotter breakup\cite{Trotter1959} between the harmonic and the physical modes, the quantum harmonic part is integrated exactly together with the Langevin thermostat for the harmonic frequencies. Indeed, the corresponding Langevin equations of motion are linear. It is then possible to exploit the mapping onto an Ornstein-Uhlenbeck process (OUP), which is exactly integrable.

To derive our algorithm, we employ the Liouvillian approach, 
that is used 
to integrate the Fokker-Planck differential equation, describing 
the evolution of the ensemble probability density 
$\rho(\bvec{p},\bvec{q},t)$\cite{Bussi2007}. 
Since the Fokker-Planck is a linear 
differential equation, this quantity 
 evolves formally as  $\rho(t+\delta t)= e^{-{\cal L} \delta t} \rho(t)$, 
with ${\cal L}$ being the Liouville operator, defined as
\begin{eqnarray}
  {\cal L} & = &  \sum_{i=1}^{3NM} \Biggl( {\cal F}_i \partial_{p_i} + p_i \partial_{q_i}  \Biggr. \nonumber \\
 & - &  \Biggl. \sum_{j=1}^{3NM} \gamma_{ij} \biggl(\partial_{p_i} p_j + k_B T M \partial_{p_i}\partial_{p_j}\biggr) \Biggr)
\label{liouvillian}
\end{eqnarray}
in mass-scaled coordinates, with $\partial_{q_i} \equiv \frac{\partial}{\partial_{q_i}}$. $ {\cal L}$ is built upon the Hamiltonian propagation, driven by the first two terms, and the Langevin thermostat, represented by the last two, deriving from the Fokker-Planck equation. For the sake of readability, $i$ and $j$ run here over all particles and beads indexed together.
In Eq.~(\ref{liouvillian}), $ {\cal F}_i \equiv f^\textrm{BO}_i + f^\textrm{harm}_i$ is the generalized force, comprising the BO and harmonic contributions, where $f^\textrm{BO}_i \equiv -\partial_{q_i}{\tilde V}$ and ${\tilde V}=\sum\limits_{j=1}^M V(q_1^{(j)},..,q_{3N}^{(j)})$.

The quantum-to-classical isomorphism Hamiltonian in Eq.~(\ref{mapping}) includes very different energy scales. 
To cope with this, we would like to split the Liouvillian in Eq.~(\ref{liouvillian}) into just two operators, one containing only the physical (BO) modes, the other depending exclusively on the fictitious (harmonic) modes. To do so, we first separate the friction matrix into two contributions:
\begin{equation}
\boldsymbol{\gamma} = \boldsymbol{\gamma}^\textrm{BO} + \boldsymbol{\gamma}^\textrm{harm}.
\label{gamma_split}
\end{equation}
We can then rewrite the total Liouvillian as the sum of two terms, ${\cal L} =  {\cal L}^\textrm{BO} + {\cal L}^\textrm{harm}$, where

\small
\begin{eqnarray}
{\cal L}^\textrm{harm} & = &  \sum_i \Biggl( f^\textrm{harm}_i \partial_{p_i} + p_i \partial_{q_i} \Biggr. \nonumber \\
                                 & - & \Biggl. \sum_j \gamma^\textrm{harm}_{ij} \biggl(\partial_{p_i} p_j + k_B T M \partial_{p_i}\partial_{p_j}\biggr) \Biggr), \label{L_harm} \\
{\cal L}^\textrm{BO} & = & \sum_i \Biggl( f^\textrm{BO}_i \partial_{p_i}\Biggr. \nonumber \\ 
                             & - & \Biggl. \sum_j \gamma^\textrm{BO}_{ij} \biggl(\partial_{p_i} p_j + k_B T M \partial_{p_i}\partial_{p_j}\biggr) \Biggr),
\label{L_BO}
\end{eqnarray}
\normalsize

\noindent in such a way that we can break up the evolution operator via Trotter factorization\cite{Trotter1959}. In order to preserve the reversibility of the process and reduce at most the time step error, we approximate the propagation using a symmetric form\cite{Tuckerman1992,Sexton}:
\begin{equation}
e^{-{\cal L}\delta t} \simeq e^{-{\cal L}^\textrm{BO} \delta t/2} e^{-{\cal L}^\textrm{harm} \delta t}  e^{-{\cal L}^\textrm{BO} \delta t/2}.
\label{secondalgo}
\end{equation}

The equations of motion corresponding to ${\cal L}^\textrm{harm}$ are those in Eq.~(\ref{ext_class_lang}), provided $\bvec{X}$ is now interpreted as a $6NM$-dimensional vector, the thermal noise $\bsym{\eta}$ lives in the $3NM$-dimensional space, and the generalized $\bsym{\hat \gamma}$ must be redefined in order to include the harmonic couplings between the beads. $\bsym{\hat \gamma}$ is now a $6NM \times 6NM$ matrix, which reads as
\begin{equation}
\hat {\boldsymbol{\gamma}} = \begin{pmatrix}
  \boldsymbol{\gamma}^\textrm{harm} & \mathbf{K} \\
 -\mathbf{I} & \mathbf{0}  
\end{pmatrix},
\label{gamma_quantum}
\end{equation}
where $\mathbf{K}$ is $3NM \times 3NM$ matrix defined as follows:
\begin{equation}
K_{ih}^{(j)(k)}=\tilde{\omega}_M^2 \delta_{ih} \left( 2 \delta^{(j)(k)} - \delta^{(j)(k-1)} - \delta^{(j)(k+1)} \right).
\label{K}  
\end{equation}
In the above definition, we have used lower indices to indicate the particle components, the upper ones (in parenthesis) indicate the bead components, while $\delta_{ij}$ ($\delta^{(i)(j)}$ for the bead indices) is the usual Kronecker delta. The $\bvec{K}$ matrix is diagonal in the particle sector, as the harmonic springs in the fictitious $H_M$ of Eq.~(\ref{mapping}) couple different replicas only for the same particle components. Cyclic conditions are implicitly applied in Eq.~(\ref{K}) to the matrix boundaries in the bead sector (i.e. $(0)=(M)$), as the polymers are necklaces.

The most notable difference of the ${\cal L}^\textrm{harm}$ equations of motion with respect to those of Eq.~(\ref{ext_class_lang}) is that $\bvec{F} = \mathbf{0}$, as the BO component of the total force has been loaded in the ${\cal L}^\textrm{BO}$ Trotter factors. Therefore, the ${\cal L}^\textrm{harm}$ equations are \emph{linear} in both $\bvec{p}$ and $\bvec{q}$, thus exactly solvable in an analytic closed form. They describe an OUP, and their algebra and properties have been exploited in the GLE framework\cite{Ceriotti2010_2}, where they are used to propagate external auxiliary variables together with the physical momenta in order to generate a colored noise with a corresponding effective memory kernel. In the present algorithm we are going to use the OUP to integrate directly the SDE for both $\bvec{p}$ and $\bvec{q}$, together with the Langevin noise, namely without further splitting the Langevin thermostat in ${\cal L}^\textrm{harm}$.

As we have done in Eq.~(\ref{pauli_expansion}) for the classical case, we expand $\bsym{\hat \gamma}$ of Eq.~(\ref{gamma_quantum}) in Pauli matrices, as follows:

\footnotesize
\begin{equation}
 \boldsymbol{\hat \gamma} = \frac{\bsym{\gamma}^\textrm{harm}}{2} \otimes \mathbf I +\frac{\bvec{K}-\bvec{I}}{2} \otimes \boldsymbol{\sigma}_x + i\frac{\bvec{K}+\bvec{I}}{2} \otimes \boldsymbol{\sigma}_y + \frac{\bsym{\gamma}^\textrm{harm}}{2} \otimes \boldsymbol{\sigma}_z.
\label{pauli_expansion_quantum}
\end{equation}
\normalsize

\noindent Then, Eq.~(\ref{pauli_expansion_quantum}) and $\left[ \mathbf{K}, \boldsymbol{\gamma}^\textrm{harm} \right] =0$ (condition fulfilled, being $\bsym{\gamma}^\textrm{harm}$ bead independent) allow one to evaluate $e^{\bsym{\hat \gamma} \delta t}$ in a closed analytic form for each (upper and lower) block component of the SDE formal solution in Eq.~(\ref{sol_ext_class_lang}). The related algebra is quite tedious and we refer the reader to the Appendix \ref{quantum_integration_scheme}. The resulting integrated equations of motions lead to the following Markov chain:
\begin{eqnarray} 
 \bvec{p}_{n+1}  &=& \bsym{\Lambda}_{1,1}   \bvec{p}_{n} +\bsym{\Lambda}_{1,2}  \bvec{q}_{n} 
+\boldsymbol \Gamma \bsym{\tilde \eta},   \label{eqdyn2_p}\\
 \bvec{q}_{n+1} & =&   \bsym{\Lambda}_{2,1}  \bvec{p}_{n} + \bsym{\Lambda}_{2,2} \bvec{q}_{n}+ 
 \boldsymbol \Theta \bsym{\tilde {\tilde \eta}},  \label{eqdyn2_q} 
\end{eqnarray}
with the integrated $6NM$-dimensional noise ($\bvec{E}_\textrm{int}$) which is
\begin{eqnarray}
\bvec{E}_\textrm{int} = 
\begin{pmatrix} 
 \boldsymbol \Gamma  \bsym{\tilde \eta} \\
  \boldsymbol \Theta \bsym{\tilde { \tilde \eta}}
\end{pmatrix} 
& = & 
\int\limits_{t_n}^{t_{n+1}} dt e^{ \boldsymbol{\hat\gamma} (t-t_{n+1})}  
\begin{pmatrix}  \bsym{\eta} \\
 0 
\end{pmatrix}. \label{defnoise}
\end{eqnarray}
The expressions for $\boldsymbol \Lambda$, $\boldsymbol \Gamma$ and $\boldsymbol \Theta$ matrices are quite complex, so they are given in Appendix \ref{quantum_integration_scheme}. 
Similarly to the classical case, we can compute the noise correlation matrix by using its definition:
\begin{equation}
\langle \bvec{E}_\textrm{int}^T \bvec{E}_\textrm{int} \rangle = \int\limits_{-\delta t}^0  dt e^{ \boldsymbol{\hat \gamma} t } 
\boldsymbol{\hat \alpha}
e^{ \boldsymbol {\hat \gamma}^\dag t },
\label{correl_alpha_tilde}
\end{equation}
where
\begin{equation}
\boldsymbol{\hat \alpha} =
\begin{pmatrix} 
 \boldsymbol \alpha & 0 \\
  0 & 0 
\end{pmatrix}.
\label{correl_alpha_hat}
\end{equation}
Once again, we refer the reader to the Appendix \ref{quantum_integration_scheme} for the analytic expression 
of the noise correlators in the quantum case. 

The propagator $e^{-{\cal L}^\textrm{harm} \delta t}$ is applied exactly, and the resulting evolution is given by the coupled equations (\ref{eqdyn2_p}) and (\ref{eqdyn2_q}) solved in the normal modes representation. $\bsym{\gamma}^\textrm{harm}$ is defined to be the optimal damping for the dynamics driven by a harmonic $H$ in its normal modes representation\cite{Mauri1994,Ceriotti2010}, namely
\begin{equation}
\gamma^{(k)}_\textrm{harm}=
\begin{cases}
2 \omega_k & \quad \textrm{if  $2 \omega_k \ge \gamma_0$} \\
\gamma_0  &  \quad  \textrm{otherwise}, 
\end{cases}
\label{gamma_PIOUMC}
\end{equation}
for each harmonic eigenmode $k=1,\ldots,M$, where 
\begin{equation}
\omega_k = 2\tilde{\omega}_M \sin \left (\frac{(k-1)\pi}{M}\right) ~~~~\textrm{for $k=1,\dots, M$}. 
\label{harm_eigen}
\end{equation}
In other words, the thermostat controlled by the friction $\gamma_0$ is mainly applied to the centroid rototranslational modes which would not be thermalized otherwise because their frequency is zero. Contrary to the other free quantum harmonic modes, the optimal damping $\gamma_0$ is not general and has to be optimized for the system under study. A reasonable value of this parameter, also present in the PILE method, can be found by performing short preliminary simulations with any accurate force field \cite{Gra2014,Babin2012,Babin2013,Babin2014}.

The equation of motion corresponding to ${\cal L}^\textrm{BO}$ is the one in Eq.~(12) of Ref.\onlinecite{Luo2014}. Indeed, only the momenta are evolved in ${\cal L}^\textrm{BO}$. Therefore, no MPC algorithm is needed, and the resulting equations are equal to those of the simple classical Langevin algorithm introduced in Ref.\onlinecite{Attaccalite2008}, restricted to $\bvec{p}$, which in the present case is a $3NM$-dimensional vector. The corresponding $\bsym{\gamma}^\textrm{BO}$ is taken block diagonal according to the bead index, and possibly bead dependent. This is very useful for QMC noisy forces, as the statistics of the QMC force covariance matrix is genuinely bead dependent, because the wave function is sampled independently for each bead (see Sec.~\ref{QMC}). In the QMC case, $\bsym{\gamma}^\textrm{BO}$ is fundamental to correct the intrinsic QMC noise affecting the BO forces. However, for deterministic BO forces, the corresponding $\bsym{\gamma}^\textrm{BO}$ can be taken equal to zero, and the corresponding time-discretized solution will result in a simple velocity update $\bvec{p}_{n+1} = \bvec{p}_n + \frac{\delta t}{2} \bvec{f}_n$. In fact, any additional damping in the BO modes will turn into a slower algorithmic diffusion, as the thermalization is already guaranteed by the Langevin thermostat integrated in ${\cal L}^\textrm{harm}$. In this limit, ${\cal L}^\textrm{BO}$ will reduce to ${\cal L}_\textrm{V}$, using the same notations as in Ref.~\onlinecite{Ceriotti2010}.

For deterministic forces, Ceriotti and coworkers\cite{Ceriotti2010} developed a different algorithm (dubbed PILE), where the main difference with respect to ours is that ${\cal L}^\textrm{harm}$ is split into ${\cal L}_0$, which propagates the motion of the harmonic oscillators in their normal modes representation, and ${\cal L}_\gamma$ (the part proportional to $\gamma_{ij}^{\rm harm}$ in Eq.~(\ref{L_harm}), ${\cal L}_0$ being the remaining part in the first line of the same equation), which corresponds to the Langevin thermostat acting on the normal modes. The total Liouvillian propagator proposed in Ref.~\onlinecite{Ceriotti2010} is:
\begin{equation}
e^{-{\cal L}\delta t} \simeq e^{-{\cal L}_{\gamma}\delta t/2}e^{-{\cal L}_V \delta t/2} e^{-{\cal L}_0\delta t}e^{-{\cal L}_V \delta t/2} e^{-{\cal L}_{\gamma}\delta t/2}.
\label{ceriottialgo}
\end{equation}
Apart from the additional Trotter breakup, in the above sequence of operators the Langevin thermostat on the normal modes is the outermost part of the Liouvillian decomposition, for a better control of the target temperature. Our algorithm in Eq.~(\ref{secondalgo}) performs better than the solution proposed in Eq.~(\ref{ceriottialgo}), since in our case ${\cal L}^\textrm{harm} ={\cal L}_0 + {\cal L}_\gamma$ corresponds to an exact propagator including both the coordinates change and the thermalization of the free quantum ring polymers. As a consequence, one Trotter factorization is saved which in principle enables us to work with a larger time step $\delta t$ or more quantum replicas $M$. 
Indeed, the Trotter break-up is obviously
exact if the two operators involved commute. The split of 
$\mathcal{L}^\textrm{harm}$ proposed by Ceriotti is a very unfortunate choice as the 
commutator of $ [\mathcal{L}_0,\mathcal{L}_{\gamma}]$ is dominated by the term 
$ \gamma_{ij}^\textrm{harm} f^\textrm{harm}_j \partial_{p_i}$ which diverges 
as $M^3$ for large $M$, as it can be easily seen by the exact expression 
of the harmonic forces and the choice of the optimal friction.   
This argument suggests that our method, involving commutators at most 
diverging as $M^2$, should have a time step error much better behaved for 
large $M$.  

These algorithmic improvements will be tested and discussed in the next section.

\subsection{\label{num}Algorithm stability with deterministic forces}

We test the robustness of the PIOUD algorithm detailed in Sec.~\ref{quant_best} by comparing it with PILE. We perform PILD simulations on the Zundel ion with almost exact deterministic forces which are simply computed by finite differences of the CCSD(T) PES provided by Huang and coworkers\cite{huang}. Since our only aim here is to test the different integration schemes, the number of MD iterations $N_\textrm{iter} = 100000$ is quite small to visit the entire phase space but sufficient to identify some possible weakness of the propagators. 

A robust PILD algorithm must be stable with both large time step $\delta t$ and large number of quantum replicas $M$. Indeed, the collective modes of a large ring polymer become very stiff, so much more difficult to control, especially if the integration time step is large. Close values of the virial $T_{M,\textrm{vir}}(\mathbf{q})= \frac{N}{2\beta} + \frac{1}{2M}\sum \limits_{i=1}^{3N}\sum \limits_{j=1}^M\left(q_i^{(j)}-\bar{q_i}\right) \partial_{q_i^{(j)}}{\tilde V}$ and the primitive $T_{M,\textrm{pri}}(\mathbf{q})=\frac{3NM}{2\beta} - \frac{1}{M}\sum \limits_{j=1}^M \frac{1}{2}\left( \frac{q_i^{(j)}-q_i^{(j-1)}}{\hbar \tau}\right)^2$  kinetic energy estimators point that we are sampling properly both positions and momenta of all particles. 

\subsubsection{Stability with respect to the number of beads $M$}
\label{stability_beads}

\begin{figure*}[!htp]
\centering 
\begin{tabular}{cc} 
      \includegraphics[width=\columnwidth]{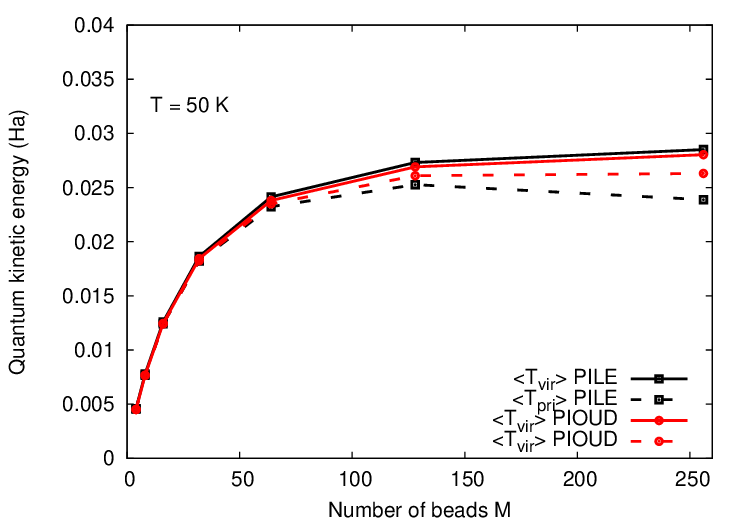} & \includegraphics[width=\columnwidth]{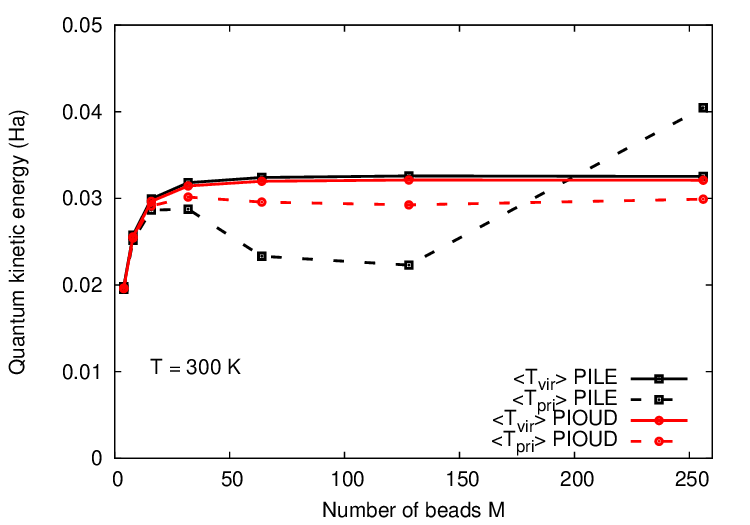}  
\end{tabular}
\caption { \label{figure:mbeads}Evolution of the quantum kinetic energy estimators $\langle T_{\textrm{vir}/\textrm{pri}}\rangle$ as a function of the number of quantum replicas $M$ at $T=50$ K (left panel) and $T=300$ K (right panel). Solid lines correspond to the virial estimator whereas the primitive estimator curves are dashed. 
We plot results for the PILE propagator (black) and the PIOUD algorithm (red). 
The time step and the friction are respectively set to $\delta t= 0.5$ fs and $\gamma_{0}=1.46 \times 10^{-3}$ a.u.} 
\end{figure*}

We report the results obtained for the PILE and PIOUD integration schemes at low temperature ($T = 50$ K) and room temperature ($T = 300$ K) in Figure \ref{figure:mbeads} as a function of the number of quantum replicas $M$. In each case, we compare the key observables $T_\textrm{vir}$ and $T_\textrm{pri}$ (other useful quantities, such as the average temperature $T$, the algorithmic diffusion constant $D$, and  the potential autocorrelation time $\tau_V$, are discussed in the Supplementary Information (SI)). We perform simulations working with an increasing number of beads $M = 4,8,16,32,64,128$ and $256$, using a reasonable but not-so-small time step $\delta t = 0.5$ fs.

The quantum kinetic energy estimators are known to be useful tools to determine the number of beads required at a given temperature to capture NQEs. In the case of the Zundel ion, $M = 128$ quantum replicas are enough to fully recover the quantum kinetic energy at low temperature, whereas only $M = 32$ beads are required at room temperature, according to Figure \ref{figure:mbeads}. We note that these values are consistent with those frequently used in the literature for this system\cite{parri_nature,Spura2015,Suzuki2013}. 

The computational efficiency and the performances of the PIOUD propagator of Eq.~(\ref{secondalgo}) and the PILE algorithm are very close to each other, regarding the average simulation temperature or the algorithmic diffusion constant (see SI for more details). This is expected because the same normal mode transformation is applied in both approaches. 

Nevertheless, when looking at the Figure \ref{figure:mbeads}, we notice that the PIOUD algorithm displays an almost perfect control of the kinetic energy operators at each temperature. On the contrary, the PILE algorithm exhibits a significant instability of the primitive energy for a large number of beads at both low and room temperature. The robustness of our approach mainly relies on a simultaneous control of both positions and velocities during the dynamics via momentum-position correlation matrices. 

In summary, the PIOUD propagator appears to be more efficient than PILE as the positions are better controlled, which enables us to work with a larger number of quantum replicas at a fixed time step $\delta t$.

\subsubsection{Stability with respect to the time step $\delta t$}
\label{stability_time}

\begin{figure}[!htp]
\centering 
      \includegraphics[width=\columnwidth]{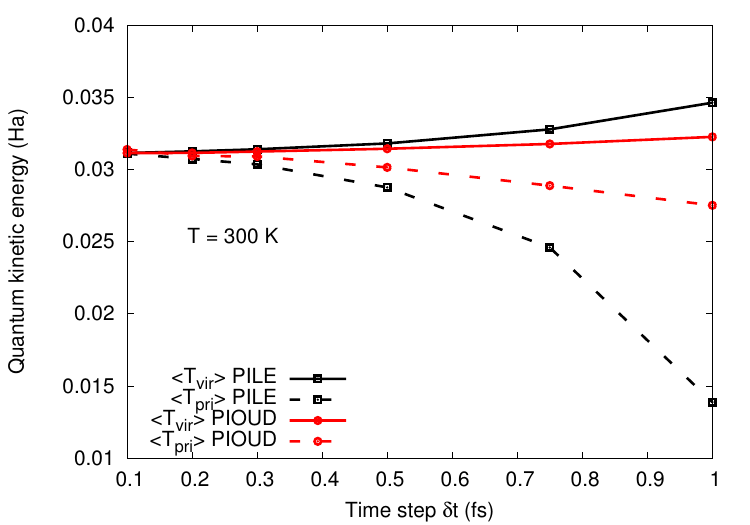} \\
\caption { \label{figure:300dt}Evolution of the quantum kinetic energy estimators $\langle T_{\textrm{vir}/\textrm{pri}} \rangle$ as a function of the time step $\delta t$ at $T=300$ K. Solid lines correspond to the virial estimator, whereas the primitive estimator curves are dashed. 
$M = 32$ quantum replicas are used and the friction is set to $\gamma_0=1.46 \times 10^{-3}$ a.u.. The color code is the same as in Figure \ref{figure:mbeads}.} 
\end{figure}

Thanks to the preliminary analysis discussed in Sec.~\ref{stability_beads}, we know how many quantum replicas should be used to generate an efficient and converged PILD simulation of the Zundel ion at a given temperature. Moreover, stability for rather large time steps is crucial in the perspective of performing PILD calculations on larger systems using a PES evaluated by accurate but computationally demanding \emph{ab initio} methods. In this Section, we report CC-PILD simulations at room temperature with $M = 32$ beads using different values of $\delta t = 0.1, 0.2, 0.3, 0.5, 0.75$ and $1$ fs. The behavior of the observables used to evaluate the algorithm efficiency are plotted in Figure \ref{figure:300dt} as a function of the time step at 300 K. 

In order to quantify the bias induced by the time step error, we check the difference $| \langle T \rangle_{M,\textrm{vir}} - \langle T \rangle_{M,\textrm{pri}} |$ which gives us direct information on the accuracy of the positions sampling. Like in the previous tests, the PIOUD algorithm shows the smallest difference thanks to a good control of the primitive energy, 
due to a better integration scheme. 
The difference between these two kinetic energy estimators is quite spectacular at room temperature. Indeed, the fluctuation-dissipation contributions in ${\cal L}^\textrm{harm}$, related to damping and random forces in the dynamics, become more important as the temperature increases, while the BO forces are not so strongly affected by thermal effects. 

The supplemental analysis based on $T$, $D$, and $\tau_V$ is reported in the SI. The algorithmic tests carried out here have been applied to a deterministic PES and exact analytic forces. We now wish to go further by applying the previous formalism to the case of stochastic PES and forces, such as the ones computed in a QMC framework, without losing computational efficiency.

\section{\label{QMC} Extension to the stochastic case: correlating the noise by QMC}

\subsection{QMC-BO surface and electron wave function}
\label{QMC-wf}

After checking the performances of the integrators with deterministic forces, we shall show that the PIOUD algorithm can be easily extended to perform  
PILD simulations in a potential energy landscape $V(\bvec{q})$ evaluated 
by VMC, namely: 
\begin{equation}
V(\mathbf{q}) = \frac{\langle \Psi_{\mathbf{q}}|H(\mathbf{q})|\Psi_{\mathbf{q}}\rangle}{\langle \Psi_{\mathbf{q}}|\Psi_{\mathbf{q}}\rangle},
\label{V_wave_function}
\end{equation}
where $|\Psi_{\mathbf{q}}\rangle$ is the QMC wave function, that minimizes the expectation value of $H(\mathbf{q})$. In the QMC framework, the evaluation of $V(\bvec{q})$ and its corresponding force estimators are intrinsically noisy.
In 
Eq.~(\ref{V_wave_function})
and in the following, we use the $\mathbf{q}$ subscript to make 
apparent
that 
the VMC wave function has an explicit 
dependence on the nuclear coordinates. Using the 
BO approximation, the PES corresponds to the expectation value of the exact Hamiltonian applied to the 
wave function $| \Psi_{\mathbf{q}} \rangle$ written here as a Jastrow Antisymmetrized Geminal Power (JAGP) product\cite{Casula2004}

\small
\begin{equation}
 \Psi_{\mathbf{q}}(\mathbf{x}_1,\dots,\mathbf{x}_{N_\mathrm{el}})  = J_{\mathbf{q}}(\mathbf{r}_1,\dots,\mathbf{r}_{N_\mathrm{el}}) \Psi_{AGP,\mathbf{q}}(\mathbf{x}_1,\dots,\mathbf{x}_{N_\mathrm{el}}).
\end{equation}
\normalsize

\noindent The set $\left\{ \mathbf{x}_i = (\mathbf{r}_i,\sigma_i) \right\}_{i=1,\dots,N_\mathrm{el}}$ represents spatial and spin coordinates of the $N_\mathrm{el}$ electrons. 

On the one hand, the bosonic Jastrow factor $J_{\mathbf{q}}$ is a function of electron-electron separation and mainly describes dynamical correlations due to charge fluctuations and Van der Waals effects. It is often written as a product of one-body, two-body and three/four-body terms $J_{\mathbf{q}} = J_{1,\mathbf{q}}J_{2,\mathbf{q}}J_{3,\mathbf{q}}$. The one body term reads

\footnotesize
\begin{eqnarray}
J_{1,\mathbf{q}}& = \exp \left ( - \sum\limits_i^{N_\mathrm{el}}\sum\limits_j^{N}(2Z_j)^{3/4}u\left((2Z_j)^{1/4}|\mathbf{r}_i - \mathbf{q}_j |\right)  \right )
\label{1body}
\end{eqnarray}
\normalsize

\noindent with $u(|\mathbf{r}-\mathbf{q}|) = \frac{1-e^{-b|\mathbf{r}-\mathbf{q}|}}{2b}$ and $b$ is a variational parameter. This form is chosen to satisfy the electron-ion Kato cusp conditions, used to deal appropriately with the 
 diverging electron-nucleus Coulomb potentials ad short distances. In the Zundel ion Hamiltonian, our benchmark system, we keep the bare Coulomb potential only for hydrogen, while for the oxygen atoms we 
replace it with
the Burkatzki-Filippi-Dolg (BFD) pseudopotential\cite{filippi_pseudo}, which is smooth at the electron-ion coalescence points. Thus, $J_{1,\mathbf{q}}$ is applied only to the hydrogen atom.
Similarly,
the electron-electron cusp conditions are dealt with by the two-body Jastrow factor 
\begin{equation}
J_{2,\mathbf{q}} = \exp \left ( \sum\limits_{i < j}^{N_\mathrm{el}}u(|\mathbf{r}_i-\mathbf{r}_j|) \right ),
\label{2body}
\end{equation}
where $u$ is a function of the same form as in Eq.~(\ref{1body}), but with a different variational parameter.  
Finally, many-body correlations are included in the remaining part of the Jastrow factor
\begin{equation}
J_{3,\mathbf{q}} = \exp \left( \sum\limits_{i<j}^{N_\mathrm{el}} \Phi_{J_{\mathbf{q}}}(\mathbf{r}_i,\mathbf{r}_{j}) \right),
\end{equation}
with 
\begin{equation}
\Phi_{J_{\mathbf{q}}}(\mathbf{r}_i,\mathbf{r}_j) = \sum\limits_{a,b}^{N}\sum\limits_{\mu,\nu}^{N_\mathrm{basis}}g_{\mu,\nu}^{a,b}\Psi_{a,\mu}^{J}(\mathbf{r}_i-\mathbf{q}_a)\Psi_{b,\nu}^{J}(\mathbf{r}_j-\mathbf{q}_b),
\end{equation}
where $N_\mathrm{basis}$ is the number of $\Psi_{a,\mu}^{J}$, the Gaussian Type Orbitals (GTOs) constituting the primitive basis of each atom $a$. The Jastrow functional form has been developed on a primitive Gaussian basis set of O(3s,2p,1d) H(2s,1p) corresponding to 213 Jastrow parameters. 

On the other hand, since our system is spin unpolarized,
the Fermionic part of the wave function is expressed as an antisymmetrized product of the geminals or pairing (AGP) functions $\Phi_{\mathbf{q}}(\mathbf{x}_i,\mathbf{x}_j)$:

\footnotesize
\begin{equation}
\Psi_{AGP,\mathbf{q}}(\mathbf{x}_1,\dots,\mathbf{x}_{N_\mathrm{el}}) = \hat{A}\left[\Phi_{\mathbf{q}}(\mathbf{x}_1,\mathbf{x}_2),\dots,\Phi_{\mathbf{q}}(\mathbf{x}_{N_\mathrm{el}-1},\mathbf{x}_{N_\mathrm{el}})\right].
\end{equation}
\normalsize

\noindent The geminals are antisymmetric functions of two electron coordinates written as a product of a spatial symmetric part and a spin singlet

\footnotesize
\begin{equation}
\Phi_{\mathbf{q}}(\mathbf{x}_i,\mathbf{x}_j) = \phi_{\mathbf{q}}(\mathbf{r}_i,\mathbf{r}_j) \frac{\delta(\sigma_i,\uparrow)\delta(\sigma_j,\downarrow) - \delta(\sigma_i,\downarrow)\delta(\sigma_j,\uparrow)}{\sqrt{2}},
\end{equation}
\normalsize

\noindent where the $\phi_{\mathbf{q}}(\mathbf{r}_i,\mathbf{r}_j)$ can be expanded either over 
the primitive basis (O(5s5p2d) and H(4s2p) for the determinantal part) or the computationally more efficient hybrid orbitals $\bar \Psi^{AGP}$, also developed on GTOs,
such that
\begin{equation}
\phi_{\mathbf{q}}(\mathbf{r}_i,\mathbf{r}_j) = \sum\limits_{a,b}^{N}\sum\limits_{\mu,\nu}^{N_\mathrm{hyb}}\lambda_{\mu,\nu}^{a,b}\bar{\Psi}_{a,\mu}^{AGP}(\mathbf{r}_i-\mathbf{q}_a)\bar{\Psi}_{b,\nu}^{AGP}(\mathbf{r}_j-\mathbf{q}_b).
\end{equation}
Indeed, one can apply here the geminal embedding scheme\cite{Sorella2015} to obtain $N_\mathrm{hyb} < N_\mathrm{basis}$ geminal embedding orbitals (GEOs), to reduce the total number of parameters, that is almost proportional to the computational cost spent for the optimization of the wave function. Here, we use the 8O 2H hybrid basis set which has been proven to be the best compromise between accuracy and computational cost, due to a limited number (571) parameters describing the determinantal part of the wave function\cite{Dagrada2014}. 
This point is very important, because in the present optimization methods\cite{Sorella2001,Casula2004,Umrigar2007},
the matrices involved in the iterative procedure have linear dimension equal to 
the number $p$ of parameters involved.  The number of QMC sampling used to  
characterize stochastically a $p\times p$ matrix should be much larger than $p$, otherwise the matrix is biased if not rank deficient, and the QMC optimization methods no longer work. Thus, it is clear that the reduction of the number of parameters is very much needed at present, because it is basically proportional to the 
computational cost of QMC optimization.

In this paper we use the optimal wave function devised by Dagrada \emph{et al.}
for the same system at zero temperature; 
further technical details concerning the variational wave function can be found in Ref.~\onlinecite{Dagrada2014}.

During the dynamics, the GTO exponents 
in both the Jastrow and the AGP part of the wave function are kept frozen to make the simulation 
stable. 
At each new ionic configuration, the wave function must be reoptimized. This is done by means of the Stochastic Reconfiguration (SR)\cite{Sorella2001,Casula2004}, or the optimization method with Hessian acceleration (SRH)\cite{Umrigar2007}. As the ionic positions are smoothly connected to those of the previous MD time step, also the wave function parameters will evolve continuously. 
After each nuclear iteration, the wave function is optimized by few (five in our simulations of the Zundel ion) SR or SRH steps on the electronic parameters  
to ensure that the system is close enough to the BO surface, before 
continuing
the propagation of the ion dynamics. Due to the continuity of the nuclear trajectories, the number of SR steps is significantly smaller than the one required for a wave function optimization from scratch.

\subsection{Bead-grouping approximation}
\label{QMC-bead-grouping}

Another advantage of our framework is represented by the local, ion-centered, basis set. We start from the observation that the most relevant dependence of the wave function on the ionic positions $\bvec{q}$ comes directly (and explicitly) from the basis set, while the dependence of the electronic variational parameters is generally weaker. This is particularly true in quantum MD calculations, where each particle is represented by an $M$-bead necklace. In the first approximation, a necklace could share the same wave function parameters, while the full $\bvec{q}$ dependence is provided by the basis set only, a dependence coming from the atomic centers -different in each bead- defining the basis set at each different (quantum) time slice. This approximation can be very effective to reduce the computational burden, because the main drawback encountered in PIMD and PILD calculations with respect to their classical counterparts is the factor-of-$M$ increase of variational parameters ($M \times p$ if no approximation is employed). 
Conversely, if the number of variational parameters is restricted to the same 
number as the classical simulation, no computational overhead is expected to work with $M>1$ in QMC. Indeed, the evaluation of the ionic forces can be 
done with a number of samples inversely proportional to $M$, as the temperature in each time slice is increased by $M$ and the Langevin equations require statistically less accurate -but nevertheless unbiased- forces,  so that increasing $M$ becomes essentially cost-free in QMC\cite{Pierleoni2006}.

When instead forces are computed with deterministic methods - e.g. DFT- 
the computational burden is necessarily proportional to $M$, and therefore 
several techniques have been recently developed  to decrease the number of evaluations of the ionic forces by about an order of magnitude without missing any significant NQE. This is done by achieving smart interpolations and groupings of different possible paths\cite{Cheng2016,Poltavsky2016} or by applying a Ring Polymer Contraction (RPC) scheme\cite{Markland2008,Markland2008_2,Fanourgakis2009}. Otherwise, the number of quantum replicas can be reduced by working with a GLE including a colored noise mimicking the quantum fluctuations of the nuclei~\cite{Ceriotti2009,Ceriotti2009_2,Ceriotti2010_2,Ceriotti2011}.

Although these methods could be effectively incorporated into our QMC framework, we are forced to explore another approach,  
because, as we have discussed, in QMC the problem is just the large  number of variational parameters, and not the large value of $M$.
We introduce therefore a method 
which takes advantage of the explicit wave function representation of the electronic problem. Indeed, each bead at each iteration has its own optimal wave function $|\Psi^{(k)}_{\mathbf{q}} \rangle$, for $k=1,\dots,M$, which minimizes the variational energy at the nuclear configuration $\bvec{q}^{(k)}$.
Consequently, we need to find the best variational parameters set $\boldsymbol{\lambda}^{(k)} = \bigl\{g_{\mu,\nu}^{a(k),b(k)},\lambda_{\mu,\nu}^{a^{(k)},b^{(k)}},b^{(k)}\bigr.$, orbital contraction coefficients,$\bigl.\ldots\bigr\}$ for each wave function. Despite the availability of efficient QMC optimization algorithms, that task is still computationally demanding and would limit the application of our method to very small systems. To overcome this major difficulty, we exploit the local nature of the Gaussian basis sets used in the expansion of both the Jastrow and AGP factors. As anticipated, we make the approximation of 
defining $N_\mathrm{groups}$ groups of neighboring beads and constraining the wave function parameters to be equal for all beads in the same group.
Since a group shares the same parameters, the corresponding energy gradients 
are then averaged over the quantum replicas constituting the group. In this way, we improve the statistics by a factor of $M/N_\mathrm{groups}$.
We obtain less noisy 
parameters even though the resulting wave function is not exactly optimized for each quantum replica. We mention that this is a controllable and systematically improvable approximation. Indeed, if one takes $N_\mathrm{groups} = M$, the electronic result is exact,  whereas $N_\mathrm{groups}=1$ constitutes the roughest approximation. In the latter case, one performs a fully quantum dynamics with almost the same statistics as the one with classical nuclei. We are going to test and validate this approximation for the Zundel ion (see Sec.~\ref{results}). In the case of the simple  hydrogen molecule the approximation with $N_\mathrm{groups}=1$ allows one to recover $90\%$ of the Zero Point Energy (ZPE).

\subsection{QMC ionic forces and noise correction}
\label{QMC-noise-correction}

In our approach, the potential energy landscape and thus the ionic forces acting on each particles are evaluated by VMC. As already done Sec.~\ref{met}, we denote the $3NM$-dimensional force acting on all the $NM$ quantum replicas by 
\begin{equation}
\mathbf{f} = - \boldsymbol{\nabla}_{\mathbf{q}}V(\mathbf{q})
\label{force}
\end{equation}
where $\boldsymbol{\nabla}_{\mathbf{q}}$ is the gradient relative to the Cartesian coordinates $\mathbf{q}$ of all the images of the nuclei. Its expression is quite complex because it contains the implicit and explicit dependence of $V(\mathbf{q})$ on both the nuclear positions $\mathbf{q}$ and the electronic parameters set $\boldsymbol{\lambda}$. Computing these forces with finite variance and in a fast way is of paramount importance to make a QMC-based MD and PIMD possible. This has become feasible, thanks to recent improvements, such as the space warp coordinate transformation (SWCT)\cite{Filippi2000}, generalized to infinitesimal ion displacements via algorithmic differentiation (AD)\cite{Sorella2010}.
Thanks to the AD, computing all components of the ionic forces is only four times more expensive than the cost of an energy calculation, which is computationally affordable.

Another issue related to the stochastic nature of this method is the control of the statistical noise introduced into the dynamics. The latter must be kept under control if we want to have an unbiased sampling of the phase space
during the propagation of the trajectory. Fortunately, the methodology described in the previous sections is particularly suited for this situation,
as it deals with the most general case, where the friction $\boldsymbol \gamma$ and covariance $\boldsymbol \alpha$ matrices are not diagonal. This corresponds to the inclusion of a correlated noise into the dynamics, in contrast to the more usual white noise case. By construction, we can now make the assumption that there exists also a QMC contribution to the covariance matrix which becomes position-time dependent\cite{Attaccalite2008}
\begin{equation}
\boldsymbol \alpha(\mathbf{q}) = \alpha_\textrm{BO} \mathbf{I} + \Delta_0 \boldsymbol{\alpha}^\textrm{QMC}(\mathbf{q}).
\label{alpha_total}
\end{equation} 
$\alpha_\textrm{BO} = 2k_BT\gamma_\textrm{BO}$ is the white noise contribution and $\Delta_0$ is a tunable parameter which is mainly set to make the covariance (and thus the friction) matrix positive definite. $\boldsymbol{\alpha}$ and $\bsym{\gamma}$ are linked by the FDT, while $\boldsymbol{\alpha}^\textrm{QMC}$ is the QMC-force covariance matrix defined by
\begin{equation}
\alpha^\textrm{QMC}_{ij} = \langle \delta f_i \delta f_j \rangle,
\label{alpha_qmc}
\end{equation}
where $\delta f_i =  f_i - \langle f_i \rangle$ is the fluctuation of the \emph{i}-th ionic force. Previous works have established that the QMC covariance matrix is roughly proportional to the dynamical matrix (it is exactly proportional for harmonic forces). Therefore, it carries information on the vibrational properties of the system\cite{Luo2014}. As a consequence, one could take advantage of an apparent drawback of the QMC approach by using the intrinsic noise to drive a dynamics in the phase space with nearly optimal sampling.

However, to fulfill the FDT, the QMC noise has to be disentangled from the total Langevin noise in the following way:
\begin{eqnarray}
\eta_i = \eta_i^\textrm{ext} + \eta_i^\textrm{QMC}
\label{noisecorrect}
\end{eqnarray}
such that $\alpha^\textrm{QMC}_{ij} = \langle \eta_i^\textrm{QMC} \eta_j^\textrm{QMC} \rangle$, namely $\eta_i^\textrm{QMC}$ is the noise already present in the $i-$th QMC force component. Therefore, by using that the external noise is independent of the QMC one, 
we obtain
\begin{equation}
\langle \eta_i^\textrm{ext} \eta_j^\textrm{ext} \rangle = \alpha_{ij} - \alpha^\textrm{QMC}_{ij}.
\label{substract}
\end{equation}
The above equation is valid at the time instant $t$. However, in the schemes devised in Secs.~\ref{class_corr}, and \ref{quant_best}, the equations of motion are discretized and integrated over a finite time step $\delta t$, and also the Langevin noise ${\tilde \eta}$ appearing in those equations is integrated over the same $\delta t$. This will slightly change the form in Eq.~(\ref{substract}), leading to the following one:
\begin{equation}
\langle {\tilde \eta}_i^\textrm{ext} {\tilde \eta}_j^\textrm{ext} \rangle = {\tilde \alpha}_{ij} - \alpha^\textrm{QMC}_{ij},
\label{noisecorrelator_integrated}
\end{equation}
where ${\tilde \alpha}$ is the integrated noise correlator matrix, reported in Eq.~(\ref{totalnoise}) for the CMPC algorithm. $\bsym{\alpha}^\textrm{QMC}$ does not need to be modified, as the expectation value of the variance-covariance QMC-force matrix in Eq.~(\ref{alpha_qmc}) is evaluated in this case during the time step $\delta t$ while sampling the electron coordinates. Since the r.h.s. of Eq.~(\ref{noisecorrelator_integrated}) is a positive definite matrix (thanks to the appropriately chosen $\Delta_0 \ge \delta t$), it defines a corrected external noise which is compatible with the solution of the Langevin dynamics in both classical and quantum cases, as long as $|\bvec{q}_{n+1}-\bvec{q}_n|$ remains small, regardless of how large the friction is. 
As already mentioned, the proposed MPC schemes yield a correlated noise also affecting the conjugate variables (i. e. the nuclear positions $\bvec{q}$). Therefore, the relation in Eq.~(\ref{noisecorrelator_integrated}) must be extended in the joint momentum-position coordinates. The full expression for the noise correction in the classical CPMC integration algorithm is detailed in Appendix \ref{noise_correction}. 

In the quantum PIOUD algorithm, the QMC noise correction must be applied exclusively in the BO step ${\cal L}^\textrm{BO}$, where only the momenta are evolved according to the QMC ionic forces. Therefore, the external noise acts only on the momentum sector, and its variance-covariance is simply given by Eq.~\ref{noisecorrelator_integrated}. Contrary to the CPMC case, there is no need to extend this relation to the position sector. In the harmonic step ${\cal L}^\textrm{harm}$, the harmonic forces are noiseless, therefore no correction is needed, and the external noise coincides with the full Langevin noise ($\eta_i^\textrm{QMC}=0$).  Its momentum and position components ($\bsym{\tilde \eta}$ and  $\bsym{\tilde {\tilde \eta}}$, respectively) are correlated according to Eqs.~\ref{noisehat1} and \ref{noisehat2}, reported in Appendix \ref{quantum_integration_scheme}.

\begin{figure}[!htp]
\centering 
\begin{tabular}{c} 
      \includegraphics[width=\columnwidth]{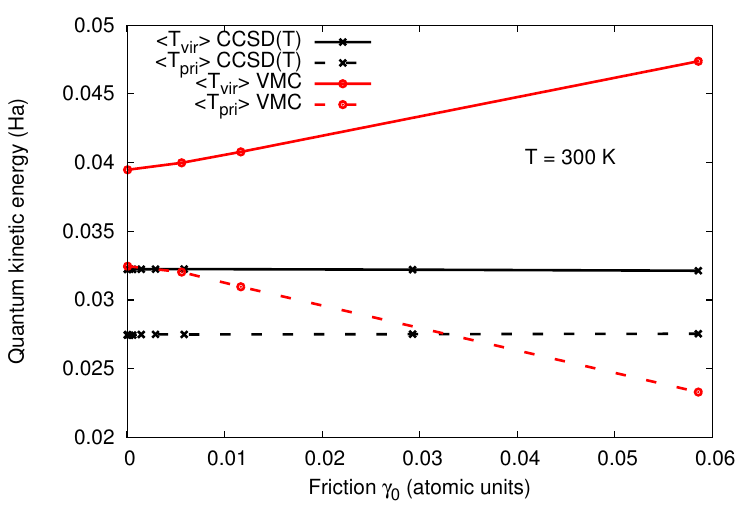} \\
      \includegraphics[width=\columnwidth]{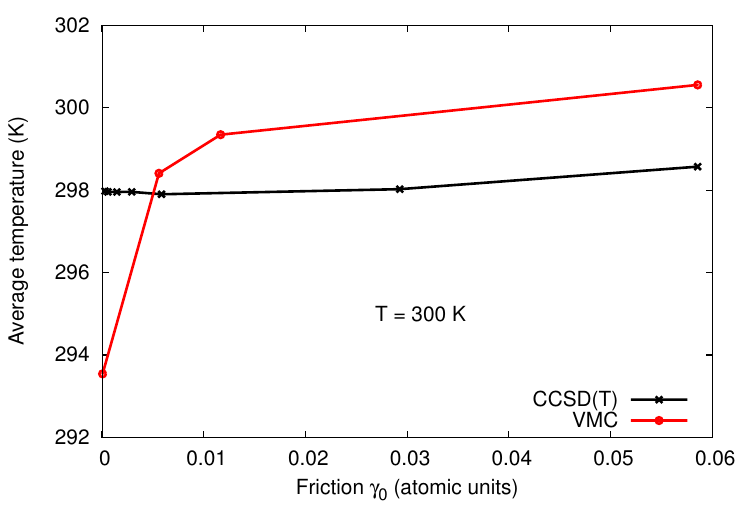} \\
      \includegraphics[width=\columnwidth]{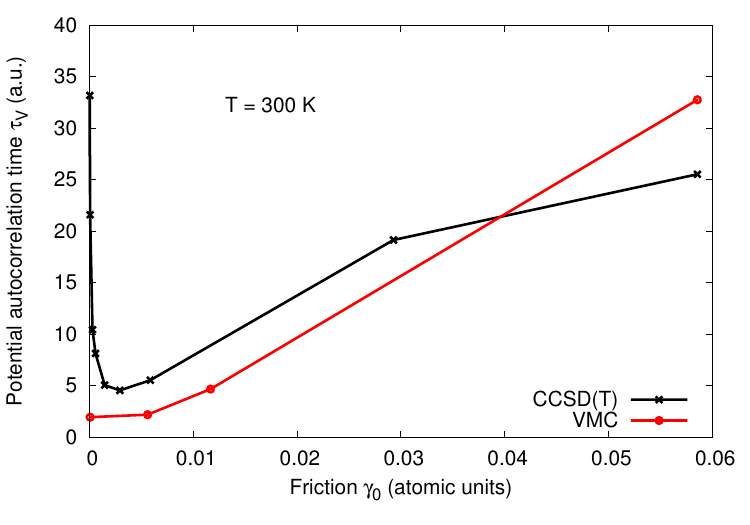} \\
\end{tabular}
\caption { \label{figure:gamma}PIOUD evolution of the quantum kinetic energy estimators $\langle T_{\textrm{vir}/\textrm{pri}}\rangle$ (top panel), the temperature $T$ (middle panel) and the potential autocorrelation time $\tau_V$ (bottom panel) as a function of the input friction $\gamma_0$. Solid lines correspond to the virial estimator of the kinetic energy whereas the primitive estimator curves are dashed. Deterministic forces are represented in black whereas the noisy QMC forces are in red. The time step and the number of quantum replicas are respectively set to $\delta t= 1$ fs and $M=32$. } 
\end{figure}

\subsection{Algorithm stability with QMC forces}
\label{QMC-stability}

In the following, we shall present some calibration runs carried out with the PIOUD algorithm in the VMC framework, which will help us set the proper simulation parameters ($\gamma_0$, $\delta t$) and show the remarkable stability of the quantum dynamics even with noisy VMC forces for large $\delta t$ and large $M$. The results are obtained by performing QMC-PILD short tests (about $8.5$ ps of dynamics) of the Zundel ion in the gas phase at room temperature ($300$ K). Hereafter, the additional PIOUD parameter $\gamma_\textrm{BO}$ is set to zero, to avoid overdamping in the BO propagator. The Langevin thermostat in the ${\cal L}^\textrm{BO}$ part plays solely the role of correcting the BO dynamics for the intrinsic VMC noise. As we have already seen, for deterministic forces ${\cal L}^\textrm{BO}$ reduces to ${\cal L}^V$, i.e. it is a simple velocity step.

For each MD step we have evaluated forces and all the energy derivatives with $\approx 3.3 \times 10^5$ MC samples, much larger than the total number of variational parameters ($p \approx 2.4\times 10^3$) of the wave function. This allows reaching an accuracy of $1.5$ mHa in the total energy per VMC energy minimization step. Between two MD steps, five QMC energy minimizations are performed with the Hessian (SRH) algorithm, in order to sample the correct BO surface. As already mentioned in Sec.~\ref{QMC-wf}, all variational parameters are evolved during the dynamics, except for the GTO exponents, which are kept frozen.

The optimal input friction $\gamma_0$ is chosen to minimize the potential autocorrelation time $\tau_V = \frac{1}{\langle V^2 \rangle - \langle V \rangle ^2}  \int \limits_{0}^{N_\textrm{iter}\delta t} dt \langle \delta V(0) \delta V(t) \rangle$ (with $\delta V = V - \langle V \rangle$ the fluctuation of the potential energy), and generate the most efficient phase space sampling during the dynamics. We shall give a general protocol to find this optimal value in the case of stochastic VMC forces, where the situation can be more complicated since the FDT is now sensitive to the QMC intrinsic noise. In our initial tests, the time step is set to 1 fs, a large value, which guarantees a quick and effective exploration of the phase space. Moreover $\gamma_\textrm{BO}=0$ and $\Delta_0 = \delta t$,  
as we discovered that the most efficient simulation is the one which minimizes the damping in the BO sector. $\Delta_0$ is taken as the minimal value which provides a positive definite $\bsym{\gamma}^\textrm{BO}$.
In the top panel of Figure \ref{figure:gamma}, we first observe that in the VMC case, the virial and the primitive kinetic energy estimators deteriorate when the applied input friction is too large ($> 0.02 \times 10^{-3}$ a.u.). In this situation, the additional QMC noise makes the coupling between the system and the thermostat too large to be fully controlled for such time step values ($\delta t = 1$ fs). On the other extreme, at very small $\gamma_0$, we see that the presence of a QMC-correlated noise tends to flatten the sharp and deep minimum of the potential autocorrelation time obtained with deterministic CCSD(T) forces (bottom panel). It indicates that, contrary to the deterministic case where there is a clear advantage to set the input friction $\gamma_0$ to its optimal value, there is more freedom to choose this parameter in VMC. Indeed, the autocorrelation time divergence shown in the CCSD(T) case for small values of $\gamma_0$ disappears with VMC forces. This is due to the implicit low-value cutoff in the $\bsym{\gamma}$ matrix provided by the intrinsic QMC noise, once the QMC-force covariance matrix is converted into an effective friction, according to Eqs.(\ref{fdcond}) and (\ref{alpha_total}). 
In practice however, the value of $\gamma_0$ cannot be too small either, in order to avoid too cold temperatures shown in the middle panel of Fig.~\ref{figure:gamma}. Consequently, we need to take the largest $\gamma_0$ before the increase of the potential autocorrelation time $\tau_V$ due to the soft-modes overdamping. This will also let us recover an acceptable target temperature (see middle panel). Therefore, $\gamma_{0} = 1.46 \times 10^{-3}$ a.u seems to be a very good compromise between autocorrelation time $\tau_V$, effective temperature, and quality of the phase space sampling revealed by the kinetic energy estimators. All subsequent runs will be performed with that value. It is interesting to note that this is optimal for both deterministic and stochastic forces.

\begin{figure}[!htp]
\centering 
\begin{tabular}{c} 
      \includegraphics[width=\columnwidth]{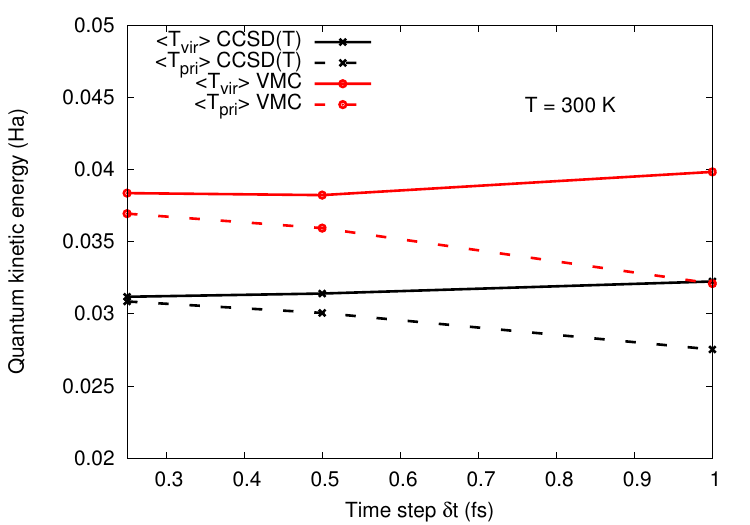} \\
            \includegraphics[width=\columnwidth]{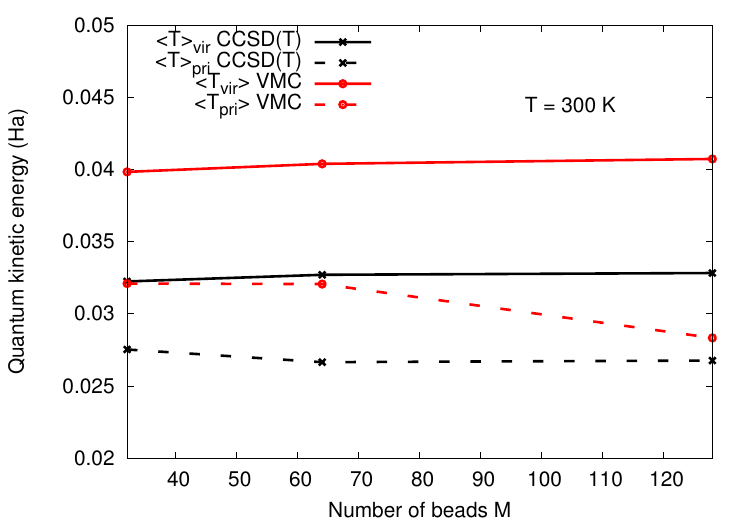} 
\end{tabular}
\caption { \label{figure:ek_all}PIOUD evolution of the quantum kinetic energy estimators $\langle T_{\textrm{vir}/\textrm{pri}}\rangle$ as a function of the time step $\delta t$ (top panel) and the number of quantum replicas $M$ (bottom panel). 
Colors and symbols are the same as in Figure \ref{figure:gamma}.
The input friction was set to $\gamma_{0}= 1.46 \times 10^{-3}$ a.u. fs. The default values of the time step and the number of quantum replicas are respectively $\delta t = 1$ fs and $M = 32$. } 
\end{figure}

Similarly to the tests performed in Sec.~\ref{num} with analytic CCSD(T) forces, we check here the robustness of our novel PIOUD algorithm with respect to the time step $\delta t$ and to the number of quantum replicas $M$ in the presence of noisy QMC forces. In the upper panel of Figure \ref{figure:ek_all}, the difference between the virial and the primitive kinetic energy estimators remains very reasonable with increasing values of the time step $\delta t$, even though the time step error is more important in the stochastic case once compared to the deterministic one. The PIOUD propagator exhibits a smaller difference between these two estimators with QMC forces than the PILE algorithm in the deterministic case. 
The superior performances of PIOUD will allow us to use large time steps $\delta t$. 

Finally, we also check the stability of the PIOUD integration scheme with increasing number of quantum replicas $M$. As we can see in the bottom panel of the Figure \ref{figure:ek_all}, the difference between the kinetic and the primitive energy estimators is well controlled up to $M = 64$ beads in the stochastic case. On the contrary, the PILE propagator already exhibits signs of instability at this value with deterministic forces (see Figure \ref{figure:mbeads}). This is a further proof of the robustness of the PIOUD integrator, which is particularly recommended when one wants to perform a PILD simulation with a large number of beads with any force field, deterministic or not.

\section{\label{results}Results}

We apply the methodology described previously in Secs.~\ref{met} and \ref{QMC} to perform QMC-based LD and PILD simulations of the Zundel ion in the gas phase at low (50 K) and room (300 K) temperature, with and without the NQE. To benchmark our results, we first carried out CMPC and PIOUD calculations using the CCSD(T) PES provided by Huang \emph{et al}\cite{huang}. In that case, the ionic forces are computed as finite differences of the potential energy with an increment $\delta q = 10^{-4}$ Bohr in the geometry. The dynamics is propagated with a time step $\delta t = 1$ fs until we obtain fully converged atomic distributions. As previously detailed, the input friction ($\gamma_\textrm{BO}$ for CMPC, $\gamma_0$ for PIOUD) is set to $1.46 \times 10^{-3}$ a.u., in order to both ensure a good thermalization of the system and minimize the potential energy autocorrelation time $\tau_V$. Consequently, that value also maximizes the diffusion of the nuclei between two subsequent LD iterations, and guarantees an efficient phase space sampling. We use 
128 beads at low temperature whereas 
32 beads are enough to fully recover NQE at room temperature, which is in agreement with Ref.~\onlinecite{Spura2015}.

The QMC-LD (CMPC) and QMC-PILD (PIOUD) simulations are performed using the wave function of Ref.~\onlinecite{Dagrada2014} as a starting point for the dynamics. 
Unless otherwise specified, all calculations are carried out by bead-grouping the VMC parameters with $N_\textrm{groups}=1$, namely with the same wave function parameters shared by all the replicas at a given iteration. The QMC statistics is the same as the one used in the stability tests (Sec.~\ref{QMC-stability}).
The equations of motion are propagated with a large time step $\delta t = 1$ fs to explore the phase space as fast as possible without doing too many evaluations of the wave function. We stopped the production run after 20 ps of dynamics, which is the minimum required to obtain converged results\cite{Sit2005}. At the end of each simulation, the average temperature, the virial and primitive kinetic energy estimators, the energy fluctuations, and the evolution of the energy gradients with respect to the electronic parameters 
are checked to ensure the reliability of the simulation. 
In order to follow the correct BO energy landscape, we require that the error made on the temperature and the kinetic energy must not exceed $5\%$, whereas the 
values of the energy gradients with respect to the wave function parameters 
must be lower than 3-4 times their standard deviation.

\begin{figure*}[!htp]
\centering 
\begin{tabular}{cc} 
      \includegraphics[width=\columnwidth]{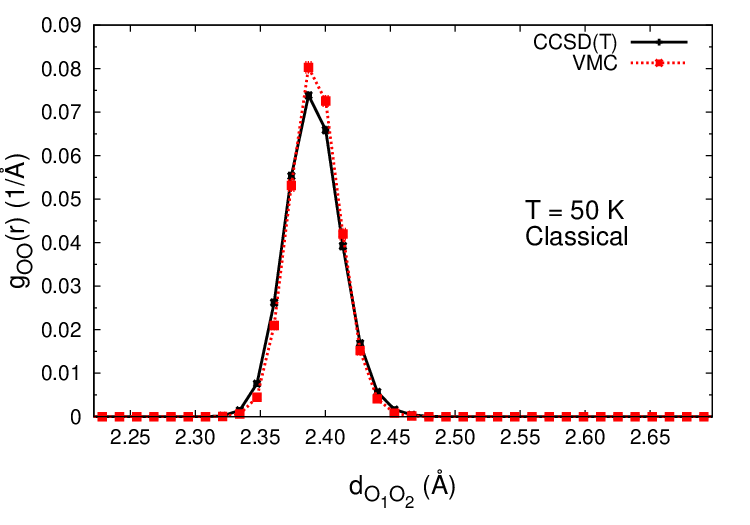} &
      \includegraphics[width=\columnwidth]{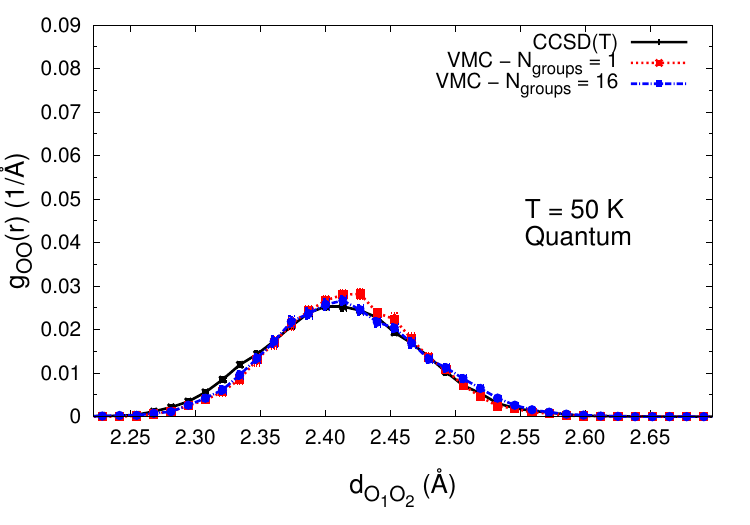} \\
      \includegraphics[width=\columnwidth]{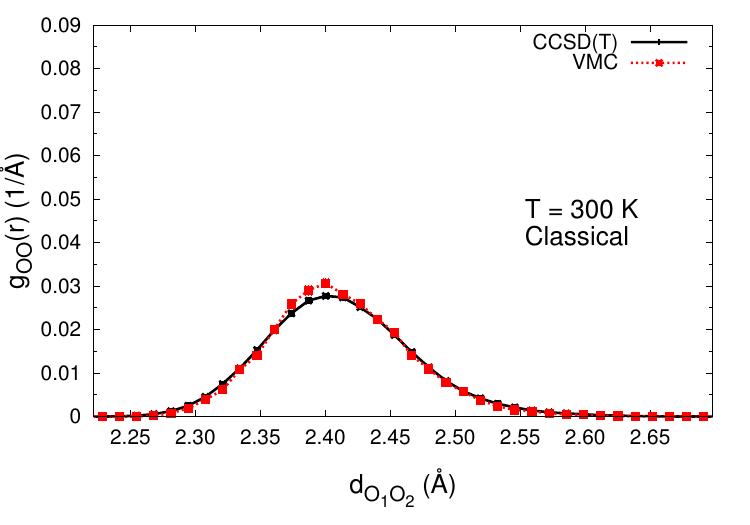} &
      \includegraphics[width=\columnwidth]{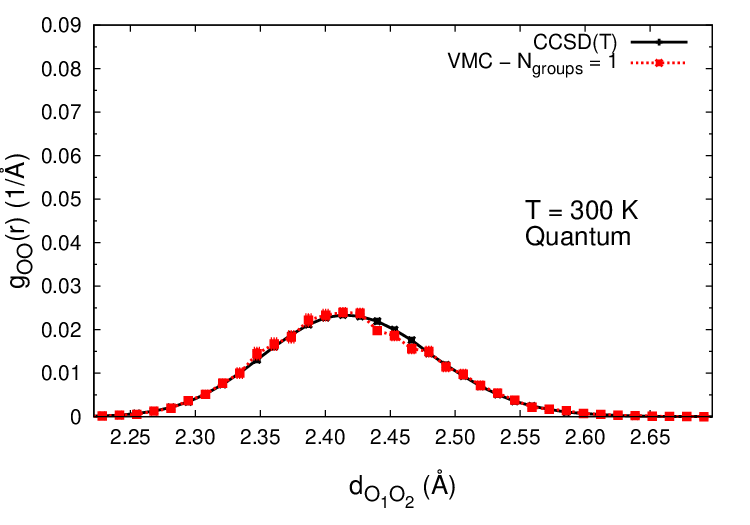} \\
\end{tabular}
\caption {\label{figure:grrOOcomp} Normalized oxygen-oxygen distributions obtained by CMPC-LD (left) and PIOUD (right) simulations at low temperature $T = 50$ K (top) and $T = 300$ K (bottom). The black curves represent the distributions obtained with analytic CCSD(T) forces, whereas the red and blue curves correspond to VMC-(PI)LD dynamics with $N_\mathrm{groups} = 1$ and $N_\mathrm{groups} = 16$ respectively. Quantum simulations are performed with $M = 128$ beads at $T = 50$ K and $M = 32$ beads at $T = 300$ K. The friction is set to $\gamma_{0} = 1.46 \times 10^{-3}$ a.u.} 
\end{figure*}

\begin{figure*}[!htp]
\centering 
\begin{tabular}{cc} 
      \includegraphics[width=\columnwidth]{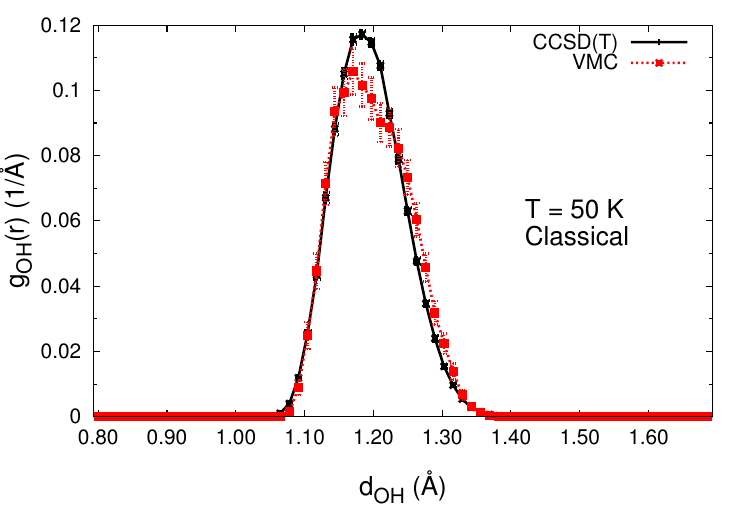} &
      \includegraphics[width=\columnwidth]{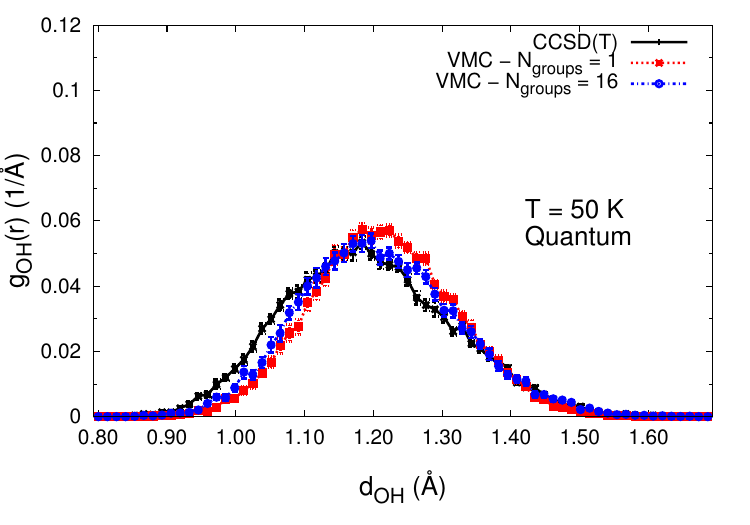} \\
      \includegraphics[width=\columnwidth]{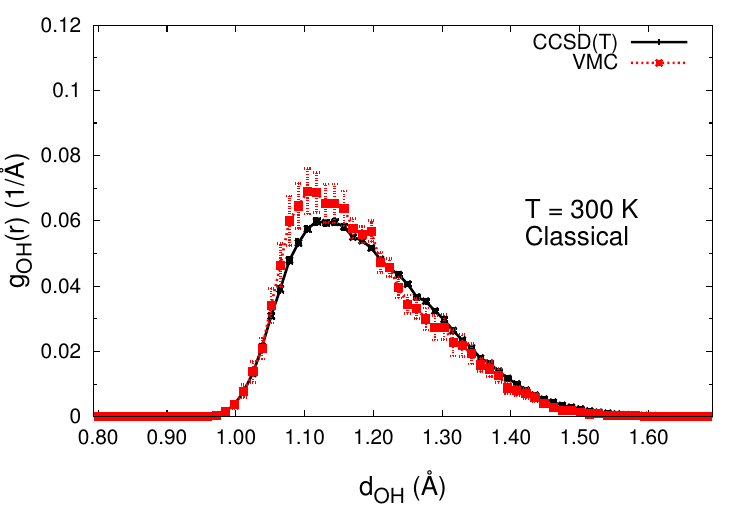} &
      \includegraphics[width=\columnwidth]{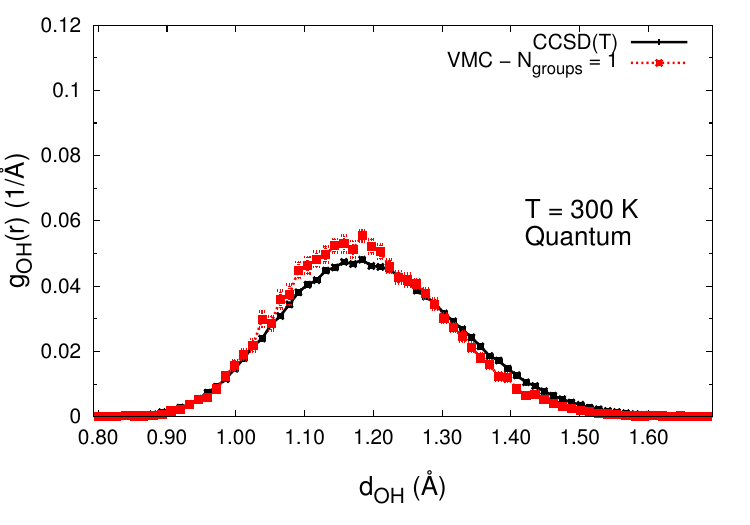} \\
\end{tabular}
\caption { \label{figure:grrOHcomp} The same as Figure \ref{figure:grrOOcomp} for the normalized oxygen-(excess) proton distributions.}
\end{figure*}

In Figures \ref{figure:grrOOcomp} and \ref{figure:grrOHcomp} we present the normalized oxygen-oxygen (g$_{\text{OO}}$) and oxygen-(excess) proton (g$_{\text{OH}}$) distributions as a function of the inter-oxygen distance. These radial distribution functions are obtained at low and room temperature with and without quantum nuclei. 
Both the VMC and benchmark CCSD(T) results are shown.

We first observe a very good overall agreement between the radial correlation functions g$_{\text{OO}}$ and g$_{\text{OH}}$ obtained with the reference CCSD(T) calculations and our VMC-based simulations. 
We notice that at 50 K, quantum effects renormalize the height of the g$_{\text{OO}}$ peak by a factor of 3, while the g$_{\text{OH}}$ distribution is renormalized by a factor of 2 with respect to its classical counterpart. The NQE have thus a huge effect in the oxygen-oxygen distribution function because the classical system is almost frozen around its zero temperature equilibrium configuration, whereas the ZPE leads to strong quantum fluctuations even at low temperature.

At 300 K, even though NQE are much less significant for the oxygen-oxygen distribution, they are still very important for the oxygen-proton correlation function, where the shapes of the distributions obtained in the classical and in the quantum case are very different. Indeed, the classical g$_{\text{OH}}$ distribution is asymmetric around the equilibrium distance, whereas the quantum correlation function is symmetric with much 
longer tails,
indicating the possibility of instantaneous proton hoppings by quantum tunneling. On the contrary, NQE are less 
dramatic
at room temperature for the g$_{\text{OO}}$ radial distribution, as expected from the greater mass of the oxygen atoms which diminishes the thermal de Broglie wavelength. 

In the classical case, the intrinsic noise present in the QMC forces tends to 
spoil
the sampling of the instantaneous oxygen-oxygen and oxygen-proton distances, which has to be accurate because the resulting distributions are extremely sharp. 
In the quantum case, instead,
the QMC noise is helpful in improving the quantum delocalization of the nuclei in the desired regions of the phase space. The phase space sampling efficiency seems
to be 
enhanced in the quantum case with respect to the classical one. The quantum results are unexpectedly easier to 
converge, and they yield radial distributions with reduced error bars compared to their classical counterparts. 

Our benchmark system is ideal also to check out the quality of the bead-grouping technique described in Sec.~\ref{QMC-bead-grouping} for the electronic parameters in \emph{ab initio} VMC-PIOUD simulations. At room temperature the bead-grouping with $N_\textrm{groups}=1$ works very well, giving results on the top of the CCSD(T) reference. At low temperature, the agreement between the CCSD(T) reference and the VMC-PIOUD results with $N_\textrm{groups}=1$ is still good, although some minor discrepancies appear in the tails of the oxygen-oxygen and oxygen-proton distributions. The strongest bias, though still quantitatively acceptable, is present in the g$_{\text{OH}}$ function, as this pair distribution is the most sensitive to quantum delocalization effects. By increasing $N_\textrm{groups}$ to 16, we improve the peak positions of both g$_{\text{OH}}$ and g$_{\text{OO}}$, and the error made in the their tails becomes almost negligible.

This can be simply interpreted by considering 
the quantum-to-classical isomorphism:
hydrogen atoms have a light mass, so the corresponding ring polymers are much more spread than the ones mimicking the quantum nuclei of oxygen. Consequently, the bead-grouping approximation on the optimal electronic parameters $\{\boldsymbol{\lambda}_\textrm{av}^{(l)}\}_{l=1,\dots,N_\mathrm{groups}}$ is more severe for hydrogen than for oxygen.
The resulting potential energy landscape is thus affected, and displays a larger curvature around its minimum, due to the energy penalty given by the non fully-optimized wave functions, being the worst for those beads which are the farthest from the centroid. This effect is apparent in the slightly shorter tails of the radial correlation function, since the corresponding ionic configurations are less visited as they have higher energies. Therefore, a compromise must be found by minimizing the total amount of CPU time spent for a simulation and the desired target accuracy on the structural and static properties of the system. While at 50 K an $N_\textrm{groups}>1$ should be chosen, at 300 K  $N_\textrm{groups}=1$ gives very accurate results. Thus, at room temperature we are able to carry out a reliable and very accurate fully-quantum dynamics of the Zundel cation in almost the same CPU time as for classical nuclei. 

\begin{figure*}[!htp]
\centering 
\begin{tabular}{cc} 
      \includegraphics[width=\columnwidth]{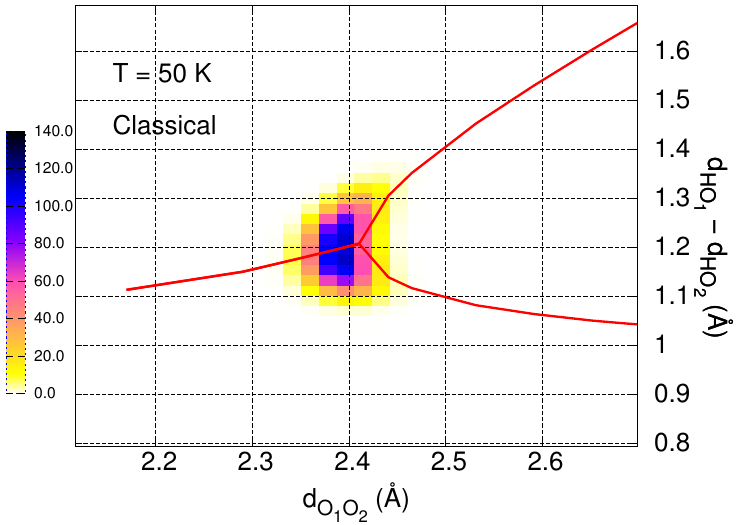} &
      \includegraphics[width=\columnwidth]{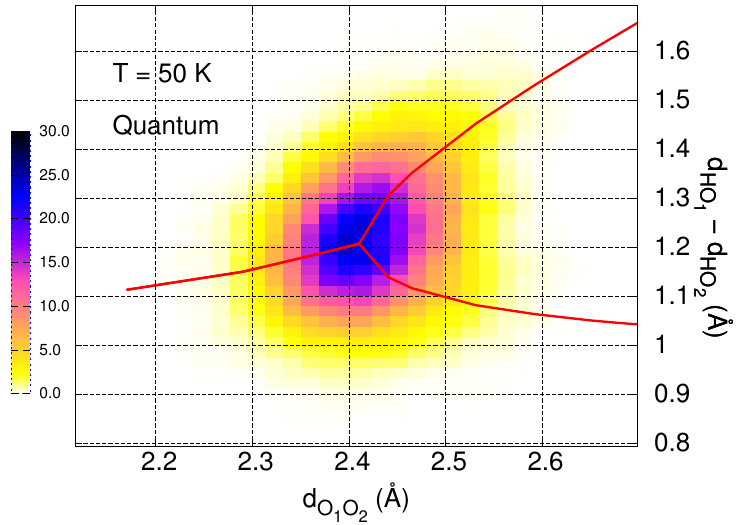} \\
      \includegraphics[width=\columnwidth]{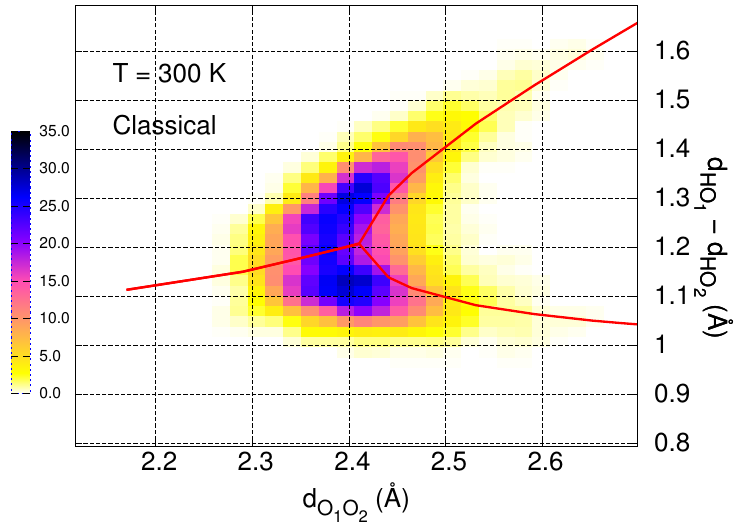} &
      \includegraphics[width=\columnwidth]{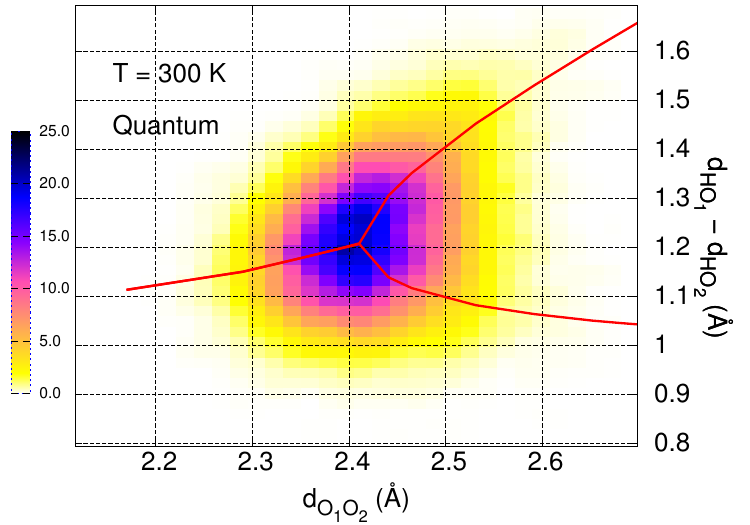} \\
\end{tabular}
\caption {\label{figure:2Dgrrcomp} Bidimensional oxygen-oxygen and oxygen-proton distributions obtained by QMC-driven CMPC-LD (left) and PIOUD (right) simulations at $T = 50$ K (top) and $T = 300$ K (bottom). The red curve corresponds to the equilibrium geometries of the Zundel ion at zero temperature obtained by CCSD(T) calculations (from Ref.~\onlinecite{Dagrada2014}). Quantum simulations are performed with $M = 128$ beads at $T = 50$ K and $M = 32$ beads at $T = 300$ K. The friction is set to $\gamma_{0} = 1.46 \times 10^{-3}$ a.u.} 
\end{figure*}

Finally, it is also interesting to quantify NQE versus thermal effects, by observing the evolution of the radial correlation functions at increasing temperature or when the quantum delocalization of the nuclei is taken into account. For classical particles, there is a clear broadening of the g$_{\text{OO}}$ and g$_{\text{OH}}$ distribution with increasing 
temperature. Indeed, the entropy is increased and the instantaneous oxygen-oxygen and oxygen-proton distances are subject to enhanced fluctuations. 
This is confirmed by examining Figure \ref{figure:2Dgrrcomp}, where we plot the oxygen-proton distance for the two oxygen sites as a function of the inter-oxygen distance. Dagrada and 
coworkers\cite{Dagrada2014} have shown that QMC techniques are in excellent agreement with the reference CCSD(T) geometries at zero temperature.
The minimum energy configuration of the Zundel cation is symmetric (C$_2$ symmetry) with the proton at the middle at the oxygen-oxygen distance. At $T = 50$ K, the classical proton remains extremely localized around the C$_2$-symmetry geometric minimum which leads to a very sharp distribution, reported 
in the upper left panels of Figures \ref{figure:grrOOcomp} and \ref{figure:grrOHcomp}. At $T = 300$ K, the system
has more thermal energy 
to visit asymmetric configurations 
with longer oxygen-oxygen distance and the excess proton sitting closer to one oxygen than to the other. This is represented by the two branches
in the lower left panel of Figure \ref{figure:2Dgrrcomp}. 
Even in this situation the symmetric configurations are more often explored than the phase regions forming the two wings.

On the contrary, thermal effects are much less important when NQE are taken into account, as apparent from 
the right panels of Figures \ref{figure:grrOOcomp}, \ref{figure:grrOHcomp} and \ref{figure:2Dgrrcomp}, which show very similar correlation functions for both temperatures.
Indeed, quantum fluctuations make the proton
able to easily jump to a neighbor oxygen site by quantum tunneling, recovering a more symmetric behavior.
This is characterized by 
longer tails 
in the g$_{\text{OH}}$ distribution function compared with the classical ones at $T = 300$ K, and by the absence of wings in Fig.~\ref{figure:2Dgrrcomp}. Moreover,  the 2D distributions  
are 
nearly the same at low and room temperature. These conclusions are in agreement with previous studies on this system\cite{Spura2015,Suzuki2013}. Some calculations performed with deterministic forces obtained from the reference CCSD(T) PES show that thermal effects start to be significant for temperatures greater than $T=900$ K. This confirms that NQE are essential to fully understand the microscopic mechanisms involving the proton dynamics in liquid water at ambient conditions.

Last but not least, we discuss the computational cost of our VMC-(PI)LD simulations. The CPU time required on HPC Marconi at CINECA (2.3 GHz 2 x 18-cores Intel Xeon E5-2697 v4 processors) to carry out a 20 ps trajectory of the Zundel ion using our VMC-(PI)LD approach is about 270k core hours. There is no significant difference between classical and quantum simulations, as long as the path integral dynamics is performed with the bead-grouping approximation with $N_\textrm{groups} = 1$. The target statistics has been detailed above and is enough to have stable and converged simulations in both electronic and ionic sectors. At the given target statistics, relaxing the bead-grouping approximation in the quantum case implies a total computational time increased by a factor $1 \le N_\textrm{groups} \le M$ with respect to the fastest case of $N_\textrm{groups} = 1$. In the perspective of applying our novel methodology to larger protonated water clusters, it is worth mentioning that for single-point calculations of equilibrium geometry at zero temperature, it has been estimated\cite{Dagrada2014} that the six-molecule complex represents a crossing point in the relative efficiency between VMC and CCSD(T) methods. By considering the excellent performances of QMC for  parallel computations, this technique is already competitive for small systems as far as the elapsed computational time is concerned. Moreover, it is obvious that a further increase in the cluster size would make the QMC approach considerably favored even in terms of total computational demand, thanks to its milder scaling with the system size.

\section{\label{conclu}Conclusions}

In this paper, we have extended the previous QMC-driven classical LD formalism (CMPC) to quantum nuclei. We have demonstrated the possibility to deal with the intrinsic QMC noise in the ionic forces evaluation by incorporating this possible source of bias into a PILD formalism. By fulfilling the FDT in a generalized framework, we correct the quantum dynamics for the QMC noise. We are thus able to perform simulations of QMC correlated electrons coupled with quantum nuclei. 

From the theoretical side, we work with joint momentum-position coordinates and exploit a Trotter breakup between the physical vibrations, set by the BO surface, and the harmonic ones, introduced by the quantum-to-classical isomorphism. This allows us to separate the dynamics driven by the BO forces from the harmonic one of the free ring polymers coupled with the Langevin thermostat. In our approach, the latter dynamics is propagated exactly, as the resulting stochastic differential equations are linear in both momenta and positions. This step corresponds to
an Ornstein-Uhlenbeck process, while in the case of QMC forces, the noise correction is performed in the former BO step. The resulting algorithm is the Path Integral Ornstein Uhlenbeck Dynamics (PIOUD).

As a first application, we took the Zundel ion, a benchmark system, and compared its quantum nuclear dynamics driven by VMC forces with the one evolved according to a deterministic PES, obtained at the accurate CCSD(T) level of theory. In the case of deterministic CCSD(T) forces, we have shown that our novel PIOUD algorithm has properties similar to the PILE scheme as far as the potential autocorrelation time $\tau_V$ and the algorithmic diffusion constant $D$ are concerned. Moreover, PIOUD features a larger stability range as a function of both time step $\delta t$ and number of quantum replicas $M$. This is encouraging, since with its improved stability, PIOUD can visit the phase space in a more efficient way, no matter whether the forces are deterministic or not. 

We further verified the stability of our approach in the stochastic case, where the PES is computed by VMC during a short nuclear dynamics. Then, we carried out extensive VMC simulations at low (50 K) and room (300 K) temperatures of the Zundel complex with both classical and quantum nuclei, by means of the CMPC and PIOUD algorithms, respectively. The outcome of the calculations in this model system suggests that even at room temperature the NQE are crucial to reach a quantitative description of proton transfer in water and aqueous systems. Indeed, NQE take over thermal effects not only at 50 K but also at 300 K, and contribute to a greater proton mobility by quantum tunneling. A very good agreement with the CCSD(T) reference calculations is found, even in the coarsest (and most effective) bead-grouping approximation, where the same electron VMC parameters are shared by all quantum replicas.

This work proves the potential of QMC methods combined with the PILD approach to perform fully quantum calculations of water-like and aqueous systems. In the perspective of performing such simulations on larger systems, our novel scheme overcomes two major scalability problems. On the one hand, it benefits from the QMC reasonable scaling with respect to the number of electrons contrary to other advanced electronic structure methods. On the other hand, using the bead-grouping technique, we are able to carry out fully \emph{ab initio} dynamics of the Zundel ion with almost the same computational cost as for classical nuclei, without deteriorating NQE. Within this approximation, the quantum simulations are paradoxically more efficient than their classical counterparts. Indeed, the nuclear observables statistics is improved, because the phase space is more efficiently explored, thanks to the quantum fluctuations included in the framework, while the QMC statistical fluctuations of the electronic part are not detrimental, as they are averaged over the beads. Consequently, this work paves the way to study, at the theoretical level, proton transfer mechanisms in more complex water clusters, starting, for instance, with the protonated water hexamer $\text{H}_{13}\text{O}_6^+$, and the larger $\text{H}_{43}\text{O}_{21}^+$ cluster, where the excess proton localization is still under debate.
Similar ideas developed in this work  for the application  
of the molecular dynamics to describe NQE, can be used  within the recently proposed ''accelerated molecular dynamics'' based on first order Langevin equation.\cite{Mazzola2016} 

\section{Associated content}

\subsection{Supporting Information}

Comparison between the PIOUD and PILE algorithms; average simulation temperature, algorithmic diffusion constant, and the potential autocorrelation time as a function of both number of beads and time step; two figures illustrating their behavior for the simulation of the Zundel ion at 300 K. This material is available free of charge via the Internet at https://pubs.acs.org/doi/suppl/10.1021/acs.jctc.7b00017.

\section*{Acknowledgements}
We thank Joel M. Bowman and Xinchuan Huang for providing us with the coupled cluster potential energy surface of $\text{H}_5\text{O}_2^+$. F.M. thanks Mario Dagrada for sharing with us the optimal $\text{H}_5\text{O}_2^+$ wave function at zero temperature, and for helpful discussions. We also acknowledge computational resources provided by the PRACE preparatory access number 2010PA2475 and PRACE projects 2015133179 and 2016133322. This work was granted access to the HPC resources of TGCC and IDRIS under the allocation 2016-096493 made by GENCI.

\appendix

\section{Noise correlators in the CMPC algorithm}
\label{cmpc_correlators}

We report the variance-covariance properties which define the time integrated noises $\bsym{\tilde \eta}$ and $\bsym{\tilde { \tilde  {\eta}}}$ for momenta and positions, respectively. Their variance-covariance can be computed by imposing the FDT (\ref{fdcond}) to be fulfilled, under the hypothesis that the $\bsym{\alpha}$ matrix is $\bvec{q}$-time independent (as we have done in Sec.~\ref{class_corr}), and by exploiting that $\left[ \boldsymbol \alpha,\boldsymbol \gamma\right]=0$. One finds:

\footnotesize
\begin{eqnarray} \label{totalnoise}
\bsym{\tilde{\alpha}}_{11} &=& \langle \bsym{\tilde \eta}^\textrm{T} \bsym{\tilde \eta} \rangle = k_B T \bsym{\gamma}^2 \coth\left(\bsym{\gamma} \frac{\delta t}{2}\right), \nonumber \\
\bsym{\tilde{\alpha}}_{22} &=& \langle  \bsym{\tilde { \tilde {\eta}}}^\textrm{T}  \bsym{\tilde { \tilde {\eta}}} \rangle = k_B T \left(2 \bsym{\Theta} - \bsym{\Gamma}^2\right) \bsym{\Theta}^{-2}, \nonumber \\
\bsym{\tilde{\alpha}}_{12} &=& \langle \bsym{{\tilde {\eta}}}^\textrm{T}  \bsym{{\tilde { \tilde {\eta}}}} \rangle = \bsym{\tilde{\alpha}}_{21}= \langle \bsym{\tilde { \tilde {\eta}}}^\textrm{T}  \bsym{{ \tilde {\eta}}} \rangle = k_B T \bsym{\gamma} \bsym{\Gamma} \bsym{\Theta}^{-1},
\end{eqnarray}
\normalsize

\noindent while the mean of $\bsym{\tilde \eta}$ and $\bsym{{\tilde { \tilde {\eta}}}} $ is zero.

\section{QMC noise correction in the CMPC algorithm}
\label{noise_correction}
We provide the noise correction which needs to be applied if one wants to include the QMC noise into the Langevin dynamics without inducing a bias on the final target temperature. Analogously to the equation (\ref{noisecorrect}), the integrated noise can be written 
\begin{eqnarray}
\tilde{\eta}_i &=& \tilde{\eta}_i^\textrm{ext} + \tilde{\eta}_i^\textrm{QMC}, \nonumber \\
\tilde{\tilde{\eta}}_i &=& \tilde{\tilde{\eta}}_i^\textrm{ext} + \tilde{\tilde{\eta}}_i^\textrm{QMC}.
\end{eqnarray}
Therefore, still assuming that the external noise  is independent from the QMC noise, we obtain very similar relations to Eq.(\ref{noisecorrelator_integrated}):
\begin{eqnarray}
\langle (\bsym{\tilde{\eta}}^\textrm{ext})^\textrm{T} \bsym{\tilde{\eta}}^\textrm{ext} \rangle &=& \bsym{\tilde{\alpha}}_{11} - \bsym{\alpha}^\textrm{QMC}, \nonumber \\
\langle (\bsym{\tilde{\tilde{\eta}}}^\textrm{ext})^\textrm{T} \bsym{\tilde{\tilde{\eta}}}^\textrm{ext} \rangle &=& \bsym{\tilde{\alpha}}_{22} - \bsym{\alpha}^\textrm{QMC}, \nonumber \\
\langle (\bsym{\tilde{\eta}}^\textrm{ext})^\textrm{T} \bsym{\tilde{\tilde{\eta}}^\textrm{ext}} \rangle &=& \bsym{\tilde{\alpha}}_{12} - \bsym{\alpha}^\textrm{QMC}, 
\end{eqnarray}
where the explicit expressions of the $\boldsymbol{\tilde{\alpha}}$ matrix components have been given in the main text in Eq.~(\ref{totalnoise}). By simply computing the square root of this $2\times 2$ matrix which is positive definite by construction, we are able to evaluate explicitly the exact integrated external noise to propagate the dynamics. 

\section{Quantum integration scheme}
\label{quantum_integration_scheme}
We detail in this section the mathematical derivation of our novel integrator described in Section \ref{PIOUD} and the explicit formulas of the $\boldsymbol \Lambda$, $\boldsymbol \Gamma$ and $\boldsymbol \Theta$ matrices. For the following algebra, it is useful to evaluate the inverse of the matrix $\boldsymbol {\hat \gamma}$
\begin{equation}
\boldsymbol{\hat \gamma^{-1}}= 
\begin{pmatrix}
\mathbf 0 & -\mathbf{I} \\
\mathbf{K^{-1}} & \mathbf{K^{-1}} \boldsymbol \gamma 
\label{invert}
\end{pmatrix},
\end{equation}
where we replaced $\bsym{\gamma}^\textrm{harm}$ of Eq.~\ref{gamma_quantum} by $\bsym{\gamma}$, for the sake of readability. 
In order to solve the differential system in Eq.~(\ref{sol_ext_class_lang}) for a generic time step $\delta t$ in the quantum case, we need to exponentiate the matrix $\boldsymbol{\hat \gamma}$. Using the fundamental assumption $ \left[ \mathbf{K}, \boldsymbol{\gamma} \right] =0$ previously justified in the paper, we consider each common eigenvector of $\mathbf{K}$ and $\boldsymbol \gamma$ that can correspond to a joint momentum and coordinate mode. In this two-fold basis, the matrix $\boldsymbol{\hat\gamma}$ is a simple $2\times2$ block matrix where as $K$ and $\gamma$ are just numbers. The block matrix $\boldsymbol{\hat \gamma}$ can be more conveniently rewritten in terms of the Pauli matrices $\boldsymbol {\sigma}_x$,$\boldsymbol {\sigma}_y$,$\boldsymbol {\sigma}_z$:
\begin{eqnarray}
 \boldsymbol{\hat \gamma} &=& \frac{\gamma}{2} \mathbf I + \mathbf x  \nonumber \\ 
\mathbf x &=&  
\begin{pmatrix}
 {\frac{\gamma}{2} } & K \\
   -1 & -{ \frac{\gamma}{2} }           
\end{pmatrix}  
\nonumber \\ 
       &=& { \frac{K-1}{2} } \boldsymbol{\sigma}_x + i { \frac{K+1}{2} } \boldsymbol{\sigma}_y + 
{ \frac{\gamma}{2} } \boldsymbol{\sigma}_z. \nonumber \\  
\end{eqnarray}
Then, the exponentiation can be straightforwardly obtained, by using standard Pauli matrices algebra
\begin{equation}
e^{ \boldsymbol{\hat \gamma }\delta t} =e^{ {\frac{\gamma \delta t}{2} } } \left\{ 
\begin{array}{cc}
\cosh( |x|\delta t) \mathbf{I} + { \frac{\mathbf x}{|x|} } \sinh( | x|\delta t ) & {\rm for ~} \gamma^2 \ge 4 K \\
\cos( |x| \delta t)  \mathbf{I}+ { \frac{\mathbf x}{|x|} } \sin( | x| \delta t ) & {\rm for ~} \gamma^2 < 4 K \\
\end{array}
\right.
\label{exponentiate}
\end{equation} 
where
$|x|= \sqrt{| \gamma^2/4-K|}$. Recombining Eq.(\ref{invert}) and Eq.(\ref{exponentiate}) one obtains:
\begin{widetext}
\begin{equation}
\boldsymbol{\hat \gamma^{-1}} ( \mathbf I - e^{-\boldsymbol{\hat \gamma} \delta t} )  = 
\begin{pmatrix}
e^{- { \frac{\gamma \delta t}{2} } } { \frac{\sinh (|x| \delta t )}{|x|} } & -(1-e^{-{ \frac{\gamma \delta t}{2} } } \cosh(|x|\delta t) ) +{\gamma  e^{- { \frac{\gamma \delta t}{2} }} \frac{\sinh( |x| \delta t )}{2 |x|} } \\
{1-e^{- { \frac{\gamma \delta t}{2} } } \frac{\cosh(|x|\delta t)}{K} }  
-{\gamma  e^{- { \frac{\gamma \delta t}{2} }} \frac{\sinh( |x| \delta t)}{2 K |x|} } &
{\gamma (1-e^{-{ \frac{\gamma \delta t}{2} }} \frac{\cosh(|x|\delta t) )}{K} }+
( 1 - \frac{\gamma^2}{2 K} ) {e^{- { \frac{\gamma \delta t}{2} }} \frac{\sinh( |x| \delta t)}{|x|} } 
\end{pmatrix} 
\end{equation}
for $\gamma^2 > 4 K$, while for $\gamma^2 \le 4 K$:
\begin{equation}
\boldsymbol{\hat \gamma^{-1}} ( \mathbf I - e^{-\boldsymbol{\hat \gamma} \delta t} )  = 
\begin{pmatrix}
e^{- { \gamma \delta t \over 2 } } { \sin (|x| \delta t ) \over |x| } & -(1-e^{-{ \gamma \delta t\over 2 } } \cos(|x|\delta t) ) +{\gamma  e^{- { \gamma \delta t\over 2 }} \sin( |x| \delta t ) \over 2 |x| } \\
{1-e^{- { \gamma \delta t \over 2 } } \cos(|x|\delta t)  \over K }  
-{\gamma  e^{- { \gamma \delta t \over 2 }} \sin( |x| \delta t) \over 2 K |x| } &
{\gamma (1-e^{- { \gamma \delta t \over 2 } } \cos(|x|\delta t) ) \over K }+
( 1 - { \gamma^2 \over 2 K }) {e^{- { \gamma \delta t\over 2 }} \sin( |x| \delta t) \over  |x| } 
\end{pmatrix} .
\end{equation}
Finally, for the case $\gamma^2 > 4 K$, we obtain:
\begin{eqnarray} 
\boldsymbol \Gamma &=&  e^{-{\gamma \over 2} \delta t } {\sinh(|x| \delta t) \over |x| }   \nonumber \\ 
\boldsymbol \Theta &=& \frac{ 1 - e^{-{\gamma \delta t \over 2 }}\cosh(|x|\delta t)}{K}-{\gamma e^{-{\gamma \delta t\over 2 }} \sinh (|x|\delta t) \over 2K|x|}  \nonumber \\
\boldsymbol \Lambda &=& e^{-{\gamma \delta t\over 2}} 
\begin{pmatrix}
 \cosh( |x|\delta t) - {\gamma \over 2} { \sinh (|x| \delta t ) \over |x| } & -K {\sinh (|x| \delta t ) \over |x| } \\
{\sinh (|x| \delta t ) \over |x| } & \cosh(|x| \delta t) + { \gamma \sinh( |x| \delta t ) \over 2 |x| } 
\end{pmatrix}, \nonumber \\
\end{eqnarray}
while for $\gamma^2 \le 4 K$, we have:
\begin{eqnarray} 
\boldsymbol \Gamma &=&  e^{-{\gamma\over 2} \delta t} {\sin(|x| \delta t) \over |x| }  \nonumber  \\ 
\boldsymbol \Theta &=& \frac{ 1 - e^{-{\gamma \delta t\over 2 }}\cos(|x|\delta t)}{K} -   {\gamma e^{-{\gamma \delta t\over 2 }} \sin (|x|\delta t ) \over 2K|x|} \nonumber  \\
\boldsymbol \Lambda &=& e^{-{\gamma \delta t\over 2}} 
\begin{pmatrix}
 \cos( |x|\delta t) - {\gamma \over 2} { \sin (|x| \delta t ) \over |x| } & -K {\sin (|x| \delta t ) \over |x| } \\
{\sin (|x| \delta t) \over |x| } & \cos(|x| \delta t) + { \gamma \sin( |x| \delta t ) \over 2 |x| } 
\end{pmatrix}. \nonumber \\
\end{eqnarray}
\end{widetext}
To evaluate the time-integrated noise covariance matrix $\bsym{\tilde{\alpha}}$ in the quantum case, defined by Eq.~(\ref{correl_alpha_tilde}), namely  
$\bsym{\tilde{\alpha}} \equiv \langle \bvec{E}_\textrm{int}^T \bvec{E}_\textrm{int} \rangle$, it is useful to introduce the following integral:
\begin{equation}
\Gamma_z=\int\limits_{-\delta t}^0 dt e^{z t}={ 1 -e^{-z \delta t} \over z }.
\end{equation}
By using Eq.~(\ref{exponentiate}), it is straightforward to carry out the integrals to obtain
\begin{eqnarray}
\bsym{\tilde{\alpha}}&=&{ \Gamma_{\gamma+2|x|} + \Gamma_{\gamma-2|x|} +2 \Gamma_\gamma \over 4} \boldsymbol{\hat {\alpha}} \nonumber  \\ 
&+&{ \mathbf{x} \boldsymbol{\hat \alpha} \mathbf{x^{\dag}} \over |x|^2 } { \Gamma_{\gamma +2 |x|}+\Gamma_{\gamma-2|x|} -2 \Gamma_\gamma \over 4}\nonumber  \\ 
&+& {\boldsymbol{\hat \alpha} \mathbf{ x}^\dag + \mathbf{x} \boldsymbol{\hat \alpha} \over |x|}  { \Gamma_{\gamma+2|x|}-\Gamma_{\gamma-2|x|} \over 4}
\label{noisehat1}
\end{eqnarray}
for $\gamma^2 \ge 4 K$, whereas for $\gamma^2 \le 4K$ we obtain a very similar formula where $|x|$ is replaced by $i|x|$:
\begin{eqnarray}
\bsym{\tilde{\alpha}}&=&{ \Gamma_{\gamma+2i |x|} + \Gamma_{\gamma-2i|x|} +2 \Gamma_\gamma \over 4} \boldsymbol{\hat \alpha} \nonumber \\  
&-&{ \mathbf{x} \boldsymbol{\hat \alpha} \mathbf{ x}^{\dag} \over |x|^2 } { \Gamma_{\gamma +2i|x|}+\Gamma_{\gamma-2i|x|} -2 \Gamma_\gamma \over 4} \nonumber \\
&-&i {\boldsymbol{\hat \alpha} \mathbf{ x}^\dag + \mathbf{ x} \boldsymbol{\hat \alpha} \over |x| } { \Gamma_{\gamma+2i|x|}-\Gamma_{\gamma-2i|x|} \over 4}.
\label{noisehat2}
\end{eqnarray}
$\boldsymbol{\hat \alpha}$ has been defined in Eq.~(\ref{correl_alpha_hat}).
In order to conclude the analytic derivation of the quantum integration scheme, we also give the explicit expression of the two matrix products appearing 
in the Eq.~(\ref{noisehat1}) and Eq.~(\ref{noisehat2}):
\begin{eqnarray}
\mathbf{ x} \boldsymbol{\hat \alpha} \mathbf{ x}^{\dag} &=& 
\begin{pmatrix}
 {\gamma^2  \alpha \over 4 } & -{\gamma \alpha \over 2} \\
 -{\gamma  \alpha \over 2}  &  \alpha 
\end{pmatrix}, \nonumber \\
\boldsymbol{\hat \alpha} \mathbf{x}^\dag + \mathbf{x} \boldsymbol{\hat \alpha}  &=&  
\begin{pmatrix}
 \gamma & - \alpha \\
- \alpha & 0 
\end{pmatrix}.
\end{eqnarray}

\bibliography{main}

\end{document}